\documentclass[aps,prl,groupedaddress]{revtex4-1}

\usepackage{graphicx}
\usepackage{dcolumn}
\usepackage{bm}
 
\def\mbi#1{\mbox{\bfseries\itshape #1}} 

\begin{document}


\title{Primordial Magnetic Field Effects on the CMB and Large Scale Structur}


\author{Dai G. Yamazaki$^{1}$}
 \email{yamazaki@asiaa.sinica.edu.tw}
\author{Kiyotomo Ichiki$^{2}$}%
\author{Toshitaka Kajino$^{3,4}$}%
\author{Grant J. Mathews$^{5}$}%

\affiliation{%
$^{1}$Institute of Astronomy and Astrophysics, Academia Sinica,
7F of Condensed Matter Sciences and Physics Department Building,
National Taiwan University.No.1, Roosevelt Rd, Sec. 4 Taipei 10617, Taiwan, R.O.C
}%
\affiliation{%
$^{2}$Department of Physics and Astrophysics, Nagoya University, Nagoya 464-8602, Japan
}%
\affiliation{%
$^{3}$
National Astronomical Observatory of Japan, Mitaka, Tokyo 181-8588, Japan
}%
\affiliation{%
$^{4}$
Department of Astronomy, Graduate School of Science, The University of Tokyo,
Hongo 7-3-1, Bunkyo-ku, Tokyo 113-0033, Japan
}%
\affiliation{%
$^{5}$Center for Astrophysics, Department of Physics, University of Notre Dame, Notre Dame, IN 46556, U.S.A.
}%


\date{December 3, 2010}

\begin{abstract}
Magnetic fields are everywhere in nature and they play an important role in every astronomical environment which involves the formation of plasma and currents.  It is natural therefore to suppose that magnetic fields could be present in the turbulent high temperature environment of the big bang.   Such a primordial magnetic field (PMF)  would be expected to manifest itself in the cosmic microwave background (CMB) temperature and  polarization anisotropies, and also in the formation of large- scale structure.  In this review we summarize the theoretical framework which we have developed 
to calculate the PMF power spectrum
to  high precision. 
Using this formulation, we summarize calculations of  the effects of a PMF which take accurate  quantitative account of the time evolution of the cut off scale. 
We review the constructed numerical program, which is without approximation, and an improvement over the approach used in a number of  previous works for studying the effect of the PMF on the cosmological perturbations.
We demonstrate how  the PMF is an important cosmological physical process on small scales. We also summarize the current  constraints on the PMF  amplitude $B_\lambda$ and the power spectral index $n_B$ which have been deduced from the available CMB observational data by using our  computational framework.
\end{abstract}


\maketitle
\section{Introduction}
Many astrophysical and cosmological phenomena in the universe are affected by magnetic fields over a broad range of scales. Indeed,  magnetic fields with a strength of $B \sim 1.0~ \mu G$ have been detected \cite{Wolfe:1992ab,Clarke:2000bz,Xu:2005rb} even on scales as large as that of galactic clusters. 
Such magnetic fields are frozen-in to the ionized baryons.  The magnetic energy density scales as $B^{2} \propto a^{-4}$ while the baryon density scale as $\rho \propto a^{-3}$.  Hence, we can relate the field strength to the cosmic density $B^3 \propto \rho^2$. Considering  that clusters of galaxies observed today would have isotropically-collapsed  relative to the background density, a cluster magnetic field  of $B \sim 1.0~\mu G$ would correspond to a primordial magnetic field (PMF) of order $\sim 1.0~n G$  at the epoch of  photon last scattering, $z \sim 1100$. 

The origin of such a cosmological primordial magnetic field has been  an area of active research by many authors. 
One expects that such a field would have a random distribution of orientations and field strength.  If  the PMF has a nearly scale invariant spectrum, an origin from vector potentials generated during the inflation epoch  is one of the best candidates
\cite{Turner:1987bw,Ratra:1991bn,Bamba:2004cu}. 
Cosmological phase transitions could also produce a PMF with a bluer spectrum
\cite{Vachaspati:1991nm,Kibble:1995aa,Ahonen:1997wh,Joyce:1997uy}.
Several authors have also discussed magnetic field generation on  smaller scales during or after the epoch of photon last scattering  ($z ^<_\sim  1100$)
\cite{Takahashi:2005nd,Hanayama:2005hd,ichiki:2006sc}.
Since each model for the generation of the PMF  involves  different length scales, the spectral index of the PMF power spectrum, $n_\mathrm{B}$ correlates with the generation models. Therefore, constraining $n_\mathrm{B}$ can lead to constraints on models for the generation of the  PMF. 

A primordial magnetic field (PMF) affects electrons and protons through the Lorentz force. Subsequently,  ionized baryons affect photons through  Thomson scattering. 
In this way the PMF affects photons indirectly in the early universe. 
A number of studies in the literature  
\cite{
Subramanian:1998fn,
Mack:2001gc,
Subramanian:2002nh,
Lewis:2004ef,
Yamazaki:2004vq,
Kahniashvili:2005xe,
Challinor:2005ye,
Dolgov:2005ti,
Gopal:2005sg,
Yamazaki:2005yd,
Kahniashvili:2006hy,
Yamazaki:2006mi,
Yamazaki:2006bq,
Yamazaki:2006ah,
Giovannini:2006kc,
Yamazaki:2007oc,
Paoletti:2008ck,
Finelli:2008xh,
Yamazaki:2008bb,
2008nuco.confE.239Y,
Sethi:2008eq,
Kojima:2008rf,
2008PhRvD..78f3012K,
Giovannini:2008aa,
Yamazaki:2009na,
2010PhRvD..81j3519Y,
2010AIPC.1269...57Y}
have analyzed effects of a  fiducial PMF of order $\sim 1~nG$ on the matter power spectrum as well as the  temperature fluctuations and polarization anisotropies of the cosmic microwave background (CMB).  
Many others 
have been, also been attempting to constrain the strength of the PMF by various means,
e.g. the non-gauusianity of the temperature fluctuations of CMB\cite{Brown:2005kr,Seshadri:2009sy,Caprini:2009vk},
Faraday rotation effects\cite{Kosowsky:1996yc,Kolatt:1997xu,Kosowsky:2004zh,Campanelli:2004pm}, 
 CMB anomalies\cite{Chen:2004nf,Bernui:2008ve,2008PhRvD..78f3012K},
and effects on  large scale structure (LSS) \cite{Sethi:2003vp,Sethi:2004pe}.
These studies have  have indicated  that the  PMF effects are  mainly manifest  on the smallest scales in the linear regime, i.e. before the formation of nonlinear structure. 
This is of interest because, there are possible discrepancies between observation and theory in the linear regime on small angular scales which could be attributable to  a PMF. However, to accurately estimate the influence  of not only a PMF but also non-linear effects, it is crucial to have the proper tools to constrain the  PMF  from cosmological observations.  The purpose of this review is to summarize the development of such tools.

We have developed methods \cite{Yamazaki:2006mi,
Yamazaki:2006bq,
Yamazaki:2006ah,
Giovannini:2006kc,
Yamazaki:2007oc,
Yamazaki:2008bb} to analyze  the  effects of a PMF on the matter and CMB power spectra.  The   parameters characterizing  the PMF have then been constrained from a fit \cite{Yamazaki:2009na} to the observational data. By simultaneously fitting  the matter and CMB contributions, we have shown that  a more comprehensive and accurate model for  the PMF effects in cosmology can be obtained.   This allows for  improved constraints on the  parameters characterizing the PMF.

In section II of this review we will introduce the basic  cosmological physics of the PMF  and in sections III and IV we will discuss its effects  on the matter power spectrum and CMB.  In section V we will discuss the results of a search for constraints on parameters of the PMF from observations of the matter power spectra and CMB. In   section IV  we will also explore the  possibilities for  new physics to be obtained from a better understanding of  the PMF, e.g. contributions of the PMF to the  gravitational wave mode (BB-mode).
 \section{The Model}
In this section we review  how to model the effects of a PMF on the cosmic fluid.   We assume   a flat Friedmann-Robertson-Walker (FRW) background cosmology for the linear perturbations and adopt a conformal synchronous gauge as in Ref.~\cite{Yamazaki:2006mi}. Hence, the  line element  is given \cite{Yamazaki:2006mi} by 
\begin{eqnarray}
ds^2=a^2(\tau)[-d\tau^2 + (\delta_{ij}+h_{ij})dx^idx^j]\label{eq_LECNG},
\end{eqnarray}
where the $x^i$ are spatial coordinates, $a(\tau)$ is the scale factor,  $h_{ij}$ is the  metric
perturbation around the background spacetime, and  $\tau$ is the conformal time defined by:
\begin{equation}
\tau \equiv \int_0^t \frac{dt'}{a(t')}~~.
\end{equation}
Here and in the following we use natural units $c=\hbar=1$ except where otherwise noted. 
 \subsection{Primordial Magnetic Field}
In order to solve the perturbed Einstein Equations it is first necessary to specify the energy momentum tensor for the magnetic field.  We can adopt  a prior upper limit on the PMF amplitude of order $\simeq 1.0-10~nG$  based upon  observations of magnetic fields on the scale of galactic  clusters as described below.
Fortunately, for a PMF of order $1.0-10$ nG at the surface of photon last scattering,  the total energy density in the  PMF is smaller than the energy density  in  the temperature fluctuations of the CMB.  Therefore, we can treat the energy density of the PMF as a first order perturbation and assume a stiff source for the time evolution. By a stiff source we mean that  all back reactions from the fluid onto the magnetic field can be discarded because these are second order perturbations. In this case, we can also assume that the conductivity of the primordial plasma is very large and that the electric field is negligible, i.e.~$E\sim 0$. This ''frozen-in'' condition is a very good and useful approximation \cite{Mack:2001gc}.

On the largest scales the time evolution of the PMF can be decoupled from its spatial dependence, i.e., $\mathbf{B}(\tau,\mathbf{x}) = \mathbf{B_0}(\mathbf{x})/a^2$.
This leads to the following simplified electromagnetic energy-momentum tensor,
\begin{eqnarray}
{T^{00}}_{[\mathrm{EM}]}(\mathbf{x},\tau)=\frac{B(\mathbf{x})^2}{8\pi a^6} \label{eq_MST_00}~~, \\
{T^{i0}}_{[\mathrm{EM}]}(\mathbf{x},\tau)={T^{0k}}_{[\mathrm{EM}]}(\mathbf{x},\tau)=0 \label{eq_MST_0s} ~~,\\
-{T^{ik}}_{[\mathrm{EM}]}(\mathbf{x},\tau)=\sigma^{ik}_\mathrm{B}=
\frac{1}{8\pi a^6}\left\{
	2B^i(\mathbf{x}) B^k(\mathbf{x}) -
\delta^{ik}B(\mathbf{x})^2
\right\}~~.
\label{eq_MST_ss}
\end{eqnarray}
 
When dealing with cosmological fluctuations, it is convenient to  work in $k$-space and denote all quantities by their Fourier transform convention
$F(\mbi{k})=\int d^3 x \exp (i\mbi{k} \cdot \mbi{x})F(\mbi{x}), $ where {\bf $k$} is a wave number.
Just as for the CMB temperature fluctuations, a PMF that is statistically homogeneous, isotropic and random, the fluctuation power spectrum can be parameterized as a
power-law $P(k) \propto k^{n_\mathrm{B}} $ \cite{Mack:2001gc,Kahniashvili:2006hy} where $n_\mathrm{B}$ is the spectral index. 

The electromagnetic stress-energy tensor in $k$ space is  given by  \cite{Yamazaki:2006mi},
\begin{eqnarray}
T^{i}_{j}(\mbi{k})_{[\mathrm{EM}]}
 &=& \frac{1}{4\pi a^{4}}\int\frac{d^3k'}{(2\pi)^3}
 	\left\{ 
 		\frac{1}{2}\delta^{i}_{j}B^{l}(\mbi{k}') B_{l}(\mbi{k}-\mbi{k}') 
 		-\frac{}{}B^{i}(\mbi{k}') B_{j}(\mbi{k}-\mbi{k}')
 	\right\}. \label{ap2} 
\end{eqnarray}

When comparing with observations, one desires a statistical measure of the fluctuations on various scales.  For this, it is best to work with a two-point correlation function for the PMF which then be defined \cite{Mack:2001gc} by
\begin{eqnarray}
\left\langle B^{i}(\mbi{k}) {B^{j}}^*(\mbi{k}')\right\rangle 
	&=&	\frac{(2\pi)^{n_B+8}}{2k_\lambda^{n+3}}
		\frac{B^2_{\lambda}}{\Gamma\left(\frac{n_B+3}{2}\right)}
		k^{n_B}P^{ij}(k)\delta(\mbi{k}-\mbi{k}'), 
		\ \ k < k_C~~,
		\label{two_point1} 
\end{eqnarray}
where $B_\lambda=|\mathbf{B}_\lambda|$ is the strength of the comoving mean magnetic field derived by smoothing over a Gaussian sphere of comoving radius $\lambda$
and $k_\lambda = 2\pi/\lambda$ 
 (with $\lambda=1$ Mpc here).   The tensor $P^{i j}(k)$ is defined by
$P^{ij}(k)=\delta^{ij}- {k{}^{i}k{}^{j}}/{k{}^2}.$
The damping scale  from radiative viscosity provides  a natural cutoff wave number $k_C$ in the PMF magnetic power spectrum. It is defined in Refs.\cite{Jedamzik:1996wp,Subramanian:1997gi,Banerjee:2004df}.

We have evolved  the PMF source power spectrum using the numerical methods described in 
Refs.~\cite{Yamazaki:2006mi,Yamazaki:2007oc}.  
Using this method, we have  quantitatively evaluated the time evolution of the cut off scale, and hence, more reliably calculated the effects of a PMF.
\subsection{Cosmological fluids, curvature fluctuatioins, and  the PMF}
\begin{figure*}[h]
\includegraphics[width=1.0\textwidth]{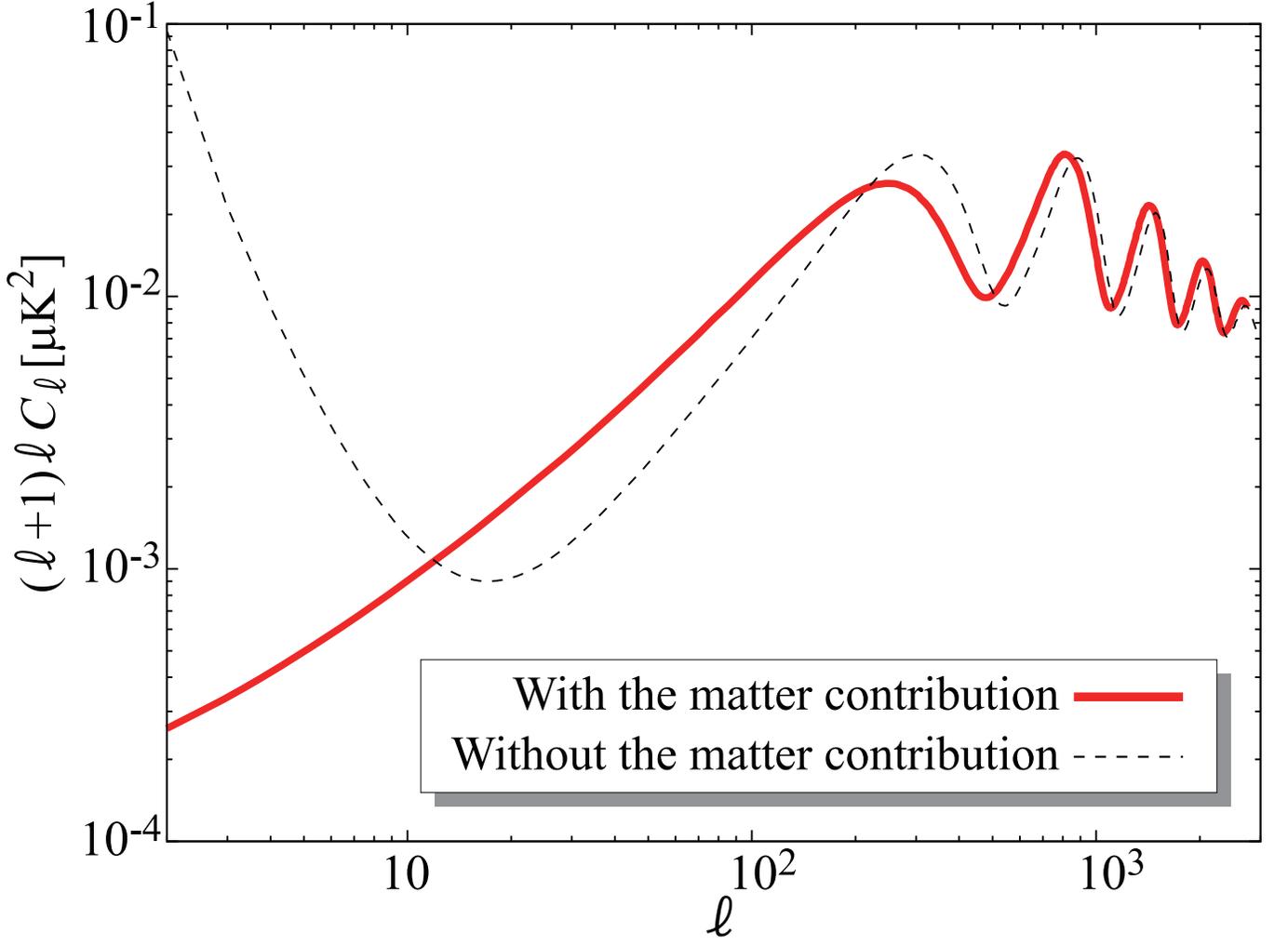}
\caption{\label{fig1} 
Comparison of CMB temperature fluctuations from the scalar mode with (red thick curve) and without (thin dashed curve) terms of $\mathcal{O}(\tau^2)$ from the matter contributions to the scale factor. For this plot we have set $B_\lambda = 1.0$ nG and $n_\mathrm{B} = -2.9$. 
}
\end{figure*}

Cosmological perturbations can be either of a scalar form (fluctuations in energy density), vector form (momentum fluctuations), or tensor form ($T^{ij}$ fluctuations).  Considering the scalar mode first, 
the ionized baryons (electrons and protons) in the early universe  are influenced by PMF Lorentz forces. Before the epoch of photon last scattering ($z > 1100$ ), the photons are immersed in a fluid of ionized baryons along with the PMF and are indirectly affected by the PMF through Thomson scattering. At the same time, the matter and radiation respond to the background curvature fluctuations $\eta$.  We have found that it is very important to carefully account for the dependence of the curvature fluctuations on the  conformal time $\tau$ up to at least second order, $\mathcal{O}(\tau^2)$. 
This is because there is a  cancellation between contributions from the primordial magnetic field and primordial radiation at first order.   Fortunately, higher  than first order $\mathcal{O}(\tau)$ terms of $\eta(\tau)$ can been  included  by considering the matter contributions to the scale factor $a$\cite{Bucher:1999re,Shaw:2009nf}, 
\begin{eqnarray}
a \simeq   \sqrt{\frac{8 \pi G \rho_{R0}}{3}}\tau
         + \frac{8 \pi G \rho_{M0}}{12}\tau^2
 = \alpha \tau + \beta \tau^2
\end{eqnarray}
where, $\rho_{R0}$ and $\rho_{M0}$ are total energy densities in radiation and matter, respectively.
A model without this matter contributions may give some values which are mathematically 
close to the correct answer for a limited range of time and length scales.  However, such models are not physically correct. 
In fact, by neglecting this coupling, the curvature perturbation $\eta$ of the scalar mode of our previous numerical estimation \cite{Yamazaki:2007oc} was too small to stabilize the numerical calculation for large scales and early times.
This problem is illustrated in Figure \ref{fig1} which shows a comparison (for $B_\lambda = 1.0$ nG and $n_\mathrm{B} = -2.9$) of the CMB temperature fluctuations in the scalar mode with and without a consideration of the matter contributions to the scale factor.  Since the previous curvature perturbation $\eta$ was too small and there was an instability in the numerical calculations for large scales and early times, the CMB temperature fluctuations for lower $\ell$  rise higher without the matter contributions to $a$ than a calculation with matter contributions.
In order to solve this problem, 
we have adopted the model of Ref.~\cite{Shaw:2009nf} for estimating the effects of the PMF on fluctuations of the scalar mode in the early universe. We have also utilized adiabatic initial conditions for the matter contributions as in Ref.~\cite{Shaw:2009nf}. 
This leads to stable numerical calculations of the curvature perturbations of the scalar mode for all length scales and times.  This is an improvement over previous numerical estimates for which the scalar curvature perturbations were too small to be stabilized in the numerical calculations for large scales and early times. Thus, our current  method allows one to obtain consistent results for all scales and times of interest.
\subsection{Correlations between the PMF and the Primary Density Field\label{s:transfer_function}}
Many authors have studied models for the origin of the PMF.  However, there is little consensus yet as to the true origin of the PMF.
Because of this, we cannot be certain of  how the PMF correlates with fluctuations of the primordial density field.
Nevertheless, one can define a parameter "$s$" to denote how much the power spectrum of the PMF correlates with  fluctuations of the primordial density field \cite{Yamazaki:2006mi,Giovannini:2006kc}. 

In the linear regime, the power spectra of density fluctuations in the baryons ($P_\mathrm{b}(k)$) and cold dark matter   ($P_\mathrm{CDM}(k)$)  in the presence of a  PMF are as follows,
\begin{eqnarray}
P_\mathrm{b}(k)&=&
	\left\langle 
		\delta_\mathrm{[b:FL]}(k)
		\delta_\mathrm{[b:FL]}^*(k)
	\right\rangle
	+
	\left\langle 
		\delta_\mathrm{[b:PMF]}(k)
		\delta_\mathrm{[b:PMF]}^*(k)
	\right\rangle 
	 \nonumber\\
	&+&
		2\left\langle 
		\delta_\mathrm{[b:FL]}(k)
		\delta_\mathrm{[b:PMF]}^*(k)
		\right\rangle ,
	 \label{CTFb}\\
P_\mathrm{CDM}(k)&=&
	\left\langle 
		\delta_\mathrm{[CDM:FL]}(k)
		\delta_\mathrm{[CDM:FL]}^*(k)
	\right\rangle
	+
	\left\langle 
		\delta_\mathrm{[CDM:PMF]}(k)
		\delta_\mathrm{[CDM:PMF]}^*(k)
	\right\rangle  \nonumber\\ 
	&+&
		2\left\langle 
		\delta_\mathrm{[CDM:FL]}(k)
		\delta_\mathrm{[CDM:PMF]}^*(k)
		\right\rangle ,
 \label{CTFCDM}
\end{eqnarray}
where the brackets denote the various two-point correlation functions as defined above, and  $\delta_\alpha$ and  $\alpha \in (\mathrm{[b:FL]}$, $\mathrm{[CDM:FL]})$
designate the baryon and CDM density fluctuations without the PMF, respectively.  While,  
$\delta_\beta$ and  $\beta \in (\mathrm{[b:PMF]}$, $\mathrm{[CDM:PMF]})$
denote the baryon and CDM density fluctuations with the PMF included (without the primary fluctuations).  
In this power spectrum we normalize the cross correlation terms with the parameter $s$,
\begin{eqnarray}
	\left\langle 
	\delta_\mathrm{[b:FL]}(k)
	\delta_\mathrm{[b:PMF]}^*(k)
	\right\rangle &\equiv& s
	\sqrt{
	        \left\langle 
		\delta_\mathrm{[b:FL]}(k)
		\delta_\mathrm{[b:FL]}^*(k)
	        \right\rangle
	        \left\langle 
		\delta_\mathrm{[b:PMF]}(k)
		\delta_\mathrm{[b:PMF]}^*(k)
	        \right\rangle
		},
\label{crossb}\\
	\left\langle 
	\delta_\mathrm{[CDM:FL]}(k)
	\delta_\mathrm{[CDM:PMF]}^*(k)
	\right\rangle 
	&\equiv& s 
	\sqrt{
	        \left\langle 
		\delta_\mathrm{[CDM:FL]}(k)
		\delta_\mathrm{[CDM:FL]}^*(k)
	        \right\rangle
	        \left\langle 
		\delta_\mathrm{[CDM:PMF]}(k)
		\delta_\mathrm{[CDM:PMF]}^*(k)
	        \right\rangle
		}~. \nonumber  \\
\label{crossCDM}
\end{eqnarray}
When $0<s\le 1$, $s=0$, or $-1\le s<0$ in eqs.(\ref{crossb}) and (\ref{crossCDM}), one has positive, vanishing, or negative correlations, respectively.

Next, the Boltzmann equation can be utilized to find the equations of for the  baryons as the follows,
\begin{eqnarray}
\dot{v}_b
	=
			-\frac{\dot{a}}{a}v_b
 			+c^2_sk^2\delta_b
 			+\frac{4\bar{\rho}_\gamma}{3\bar{\rho}_b}
 			an_e\sigma_T(v_{\gamma}-v_b)
			+k^2\frac{\rho_{\gamma 0}}{a\rho_{b 0}} \frac{\Pi_{\mathrm{[EM:S]}}(\mbi{k})}{4\pi \rho_{\gamma 0}}~,
			\label{eq:baryon_v} 
\end{eqnarray} 
where $v_b$ and $v_\gamma$ are the baryon and photon velocity perturbations,
$\delta_b$ and $\delta_\gamma$ are the baryon and photon density perturbations,
$c_s$ is the sound speed, 
$n_e$ is the free electron density, $\sigma_T$ is the Thomson scattering cross section, 
$\rho_{b 0}$ and $\rho_{\gamma 0}$ are baryon and photon densities,
and $\Pi_{\mathrm{[EM:S]}}(\mbi{k})$ is the square root of the power spectrum function for the Lorentz force 
(given in \cite{Yamazaki:2006mi}).

The Lorenz force term in Eq.(\ref{eq:baryon_v}) can be divided into two terms,
 the magnetic pressure and the tension.  
By comparing those terms, 
one can decide 
which of them is dominant in the Lorenz force.
There is, however,  no information as to whether  the magnetic pressure or the tension is dominant, and whether the direction of the forces from them are the same or different.
Nevertheless,  such information  should be taken into account when it can be determined from a model.
To see how these pieces arise, we summarize the derivation of the scalar Lorentz force term as given in Ref.~\cite{Yamazaki:2006mi}.  Analogous derivations of the vector and tensor two-point correlation functions can be found in Ref.~\cite{Yamazaki:2007oc}.  One can first rewrite the two-point correlation function for the  scalar part of the electromagnetic energy tensor in $k$ space as,
\begin{eqnarray}
\langle 
	T_{[\mathrm{EM:S}]}(\mbi{k})
	T^*_{[\mathrm{EM:S}]}(\mbi{k}')
\rangle
&=& \hat{k}_i\hat{k}^{j}\hat{k}_l\hat{k}^m
\langle
 	T^{i}_{j}{}_{[\mathrm{EM}]}(\mbi{k})
 	T^{*l}_{m}{}_{[\mathrm{EM}]}(\mbi{k}')
\rangle\nonumber\\
&=&
(2\pi)^3|\Pi_{[\mathrm{EM:S}]}(\mbi{k})|^2\delta(\mbi{k}-\mbi{k}')~~,
\label{eq:TPCFofLF}
\end{eqnarray}
where $\Pi_{[\mathrm{EM:S}]}(\mbi{k})$ is  the  power spectrum of the Lorenz force that we wish to analyze. 

One can next  decompose the electromagnetic stress-energy tensor in $k$ space into two
parts as follows,
\begin{eqnarray}
T^{i}_{j}(\mbi{k})_{[\mathrm{EM}]}
	&=&T^{i}_{j}(\mbi{k})_{[\mathrm{EM:1}]}
	-T^{i}_{j}(\mbi{k})_{[\mathrm{EM:2}]}
	\label{ap3}\\
T^{i}_{j}(\mbi{k})_{[\mathrm{EM:}1]}
	&=& \frac{1}{4\pi a^{4}}\int\frac{d^3k'}{(2\pi)^3}
 	\frac{1}{2}\delta^{i}_{j}B^{l}(\mbi{k}') B_{l}(\mbi{k}-\mbi{k}') \label{ap4} \\
T^{i}_{j}(\mbi{k})_{[\mathrm{EM:}2]}
 	&=& \frac{1}{4\pi a^{4}}\int\frac{d^3k'}{(2\pi)^3}
B^{i}(\mbi{k}') B_{j}(\mbi{k}-\mbi{k}') \label{ap5}.
\end{eqnarray}
Correspondingly, one can define \cite{Yamazaki:2006mi}
$T_\mathrm{[EM:S1]}=\hat{k}_i\hat{k}^j T^{i}_{j}{}_\mathrm{[EM:1]}$ and 
$T_\mathrm{[EM:S2]}=\hat{k}_i\hat{k}^j T^{i}_{j}{}_\mathrm{[EM:2]}$. 
Using Eq.(\ref{ap3} - \ref{ap5}),  one can finally rewrite the two point correlation function for the scalar part of the Lorenz force  as,
\begin{eqnarray}
\langle
T(\mbi{k})_\mathrm{[EM:S]}T^*(\mbi{k}')_\mathrm{[EM:S]}
\rangle
&=&
\langle
(T(\mbi{k})_\mathrm{[EM:S1]}-T(\mbi{k})_\mathrm{[EM:S2]})
(T^*(\mbi{k}')_\mathrm{[EM:S1]}-T^*(\mbi{k}')_\mathrm{[EM:S2]})
\rangle  
\nonumber\\
&=&
\langle
T(\mbi{k})_\mathrm{[EM:S1]}T^*(\mbi{k}')_\mathrm{[EM:S1]}
\rangle  
-\langle
T(\mbi{k})_\mathrm{[EM:S1]}T^*(\mbi{k}')_\mathrm{[EM:S2]}
\rangle  
\nonumber\\
&-&
\langle
T(\mbi{k})_\mathrm{[EM:S2]}T^*(\mbi{k}')_\mathrm{[EM:S1]}
\rangle  
+\langle
T(\mbi{k})_\mathrm{[EM:S2]}T^*(\mbi{k}')_\mathrm{[EM:S2]}
\rangle. \label{ap6}  
\end{eqnarray}
The first term on the R.H.S. can be identified as the magnetic pressure while the fourth term is the magnetic tension.  The second and third terms are the cross correlation.  A key part of the analysis of the effects of the PMF on the cosmic fluid is to understand the relative roles of these terms.

\begin{figure*}[h]
\includegraphics[width=1.0\textwidth]{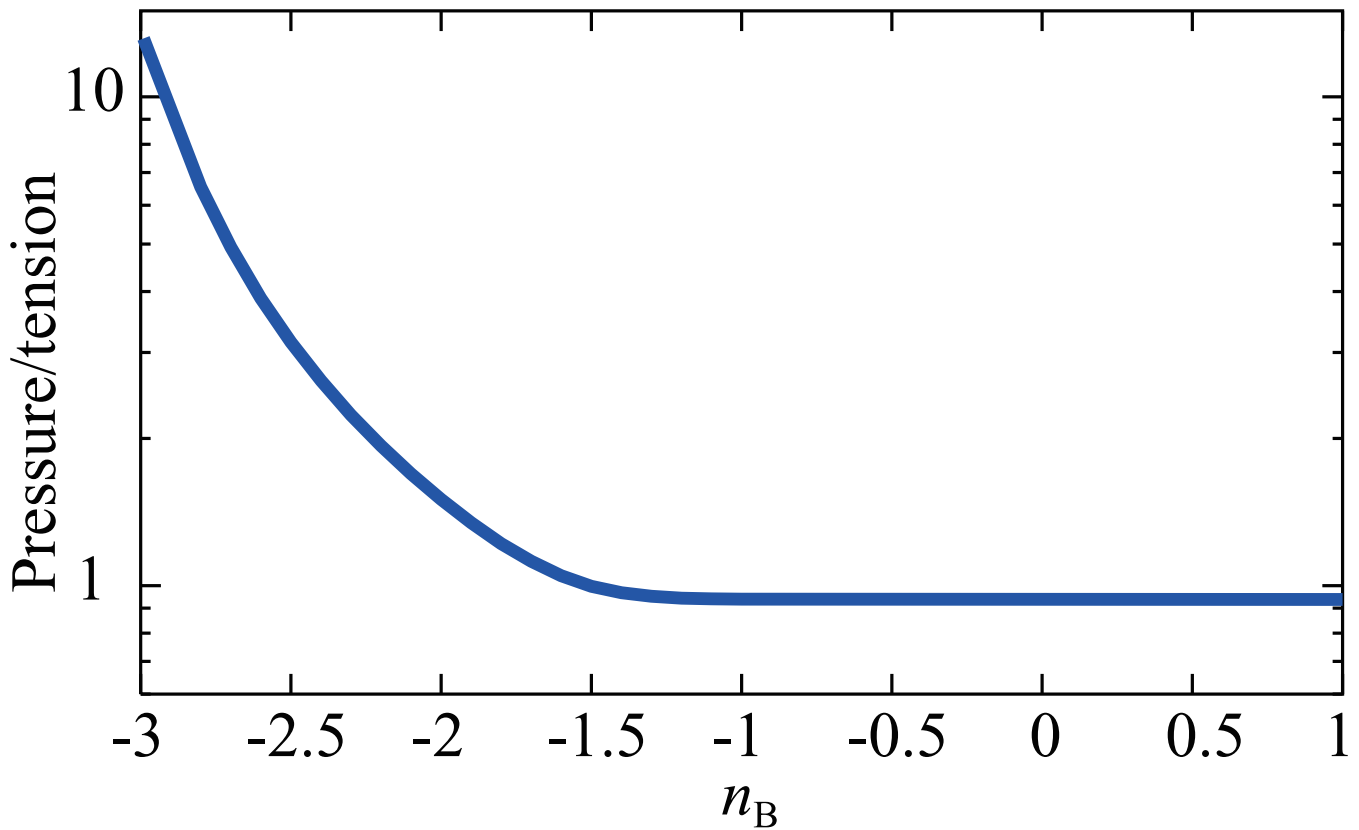}
\caption{\label{fig2new}
Ratio of stocahstic PMF pressure and tension sources. 
For illustration the cut off scale $k_C$ was fixed to $k_c = 10$ Mpc$^{-1}$.
The pressure dominates for $n_\mathrm{B} < -1.5$,
and the magnetic tension dominates for $n_\mathrm{B} > -1.5$.
}
\end{figure*}

Fig.~\ref{fig2new} shows the ratio of the magnetic pressure to  the tension as in the Lorentz force term \cite{Yamazaki:2006mi}  as a function of the spectral index $n_B$.  
The pressure  dominates when $n_\mathrm{B} < -1.5$,
while the tension is slightly larger  when $n_\mathrm{B} > -1.5$.

The sign of the cross correlation terms depends upon the relative signs of the pressure and tension.  In order to determine the relative signs of these two terms, however,   one must specify a model for the generation of the PMF.
There is as of yet no consensus model.  Therefore, we  decompose the factors into various possible combinations, i.e. 
\begin{eqnarray}
s=s_\mathrm{[LF]}\times s_\mathrm{[DF]},
\end{eqnarray}
where
\begin{eqnarray}
s_\mathrm{[LF]}=
\left\{
		\begin{array}{rl}
			-1, & n < -1.5\ ~\mathrm{(I)},\\
			-1, & n > -1.5\ ~\mathrm{(II)}, \\
			1, & n > -1.5\ \   ~\mathrm{(III)},
		\end{array}
\right.
	\label{eq:SLF}
\end{eqnarray}
and
\begin{eqnarray}
		\begin{array}{rlccc}
			0 & < & s_\mathrm{[DF]} &\le& 1\ ~\mathrm{(i)},\\
			  &   & s_\mathrm{[DF]} & = & 0\ ~\mathrm{(ii)},\\
			-1&\le& s_\mathrm{[DF]} & < & 0\ ~\mathrm{(iii)}.
		\end{array}\nonumber
\end{eqnarray}
In the different regimes,  $s_\mathrm{[LF]}$ represents either:
(I) the pressure dominated case;
(II) the tension dominated case, where the  magnetic field pressure and tension forces act in the same direction; or 
(III) the tension dominated case, where the magnetic field pressure and tension forces act in the opposite direction.
On the other hand, $s_\mathrm{[DF]}$ represents either:
(i) a positive correlation between  
the matter and PMF distributions;  (ii) no correlation;
or (iii) a negative correlation.
Thus, if $s<0$, 
the  matter and PMF distributions could  be correlated positively ($s_\mathrm{[DF]}>0$) 
and  the PMF pressure dominates in the Lorentz term (for $n < -1.5$).  
Another possibility is that 
the matter and PMF distributions negatively correlate ($s_\mathrm{[DF]}<0$) 
and the PMF tension dominates in the Lorentz term (for $n > -1.5$) and 
the tension acts on the density field in the same direction as the magnetic field pressure.
Yet another possibility is that 
the PMF tension dominates in the Lorentz term ($n > -1.5$),
but the tension acts on the density field in the opposite direction from  the pressure force.
In these cases the PMF effects act like a gas pressure to oppose the gravitational collapse and 
therefore causes the density perturbations to more slowly evolve.

On the other hand, 
if $s>0$, 
the matter and PMF distributions could positively correlate ($s_\mathrm{[DF]}>0$) 
and 
the PMF tension dominates in the Lorentz term ($n > -1.5$) while
the tension acts on the density field in the opposite direction from the pressure force.
Alternatively,  
 the matter and PMF distributions could negatively correlate
($s_\mathrm{[DF]}<0$)  
and 
the PMF pressure dominate in the Lorentz term ($n < -1.5$). 
In these cases the Lorentz force from the PMF accelerates the gravitational collapse.
After decoupling, $\delta$ does not
oscillate and the perturbation evolution is straightforward for all of the above cases. 
\section{Effects of a PMF on the Matter Power Spectrum}
In this section we review the effects of the PMF on fluctuations in the matter density on cosmological scales (see \cite{Yamazaki:2006mi} for details). 
\subsection{Effects of a PMF on the Matter Density Field }
\begin{figure*}[h]
\includegraphics[width=1.0\textwidth]{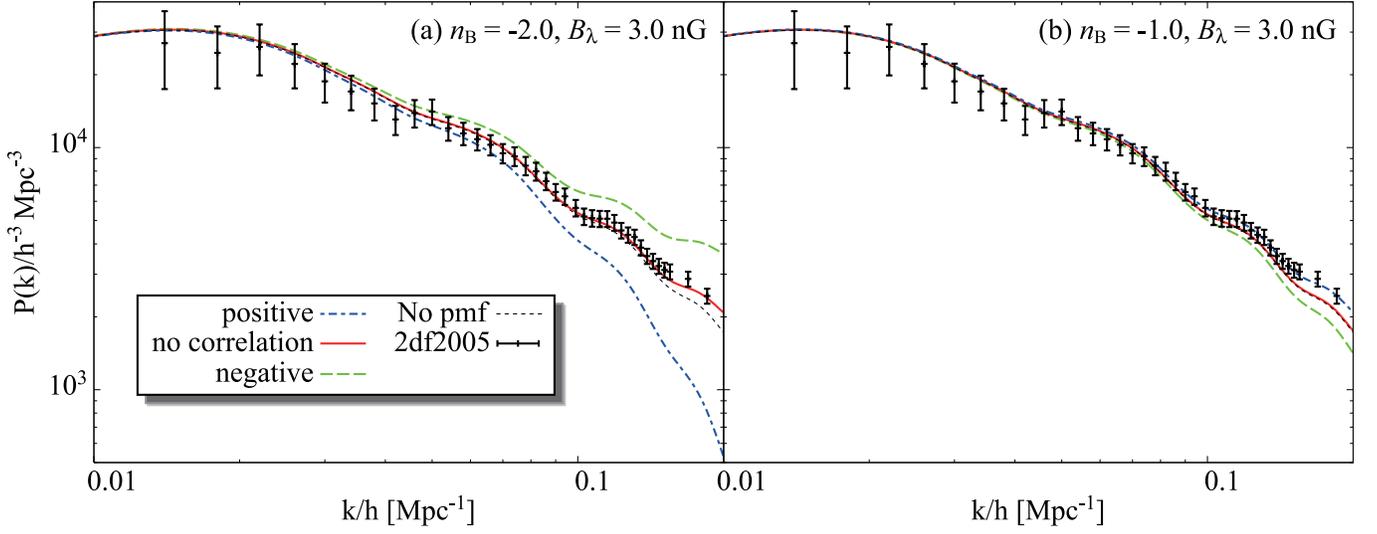}
\caption{\label{fig2} 
Effects on the matter power spectra of the PMF.
The left and right panels are for $n_\mathrm{B} = -2.0$ and $-1.0$.
The green-dashed, red-bold, and blue-dotted-dashed curves are the case of negative($s_\mathrm{[DF]}=-1$), no ($s_\mathrm{[DF]}=0$) and positive($s_\mathrm{[DF]}=1$) correlations between the matter and PMF, respectively.
A Black-dotted curve is the matter power spectrum without a PMF and the dots with error bars show the 2df results\cite{Cole:2005sx} 
}
\end{figure*}

Figure.~\ref{fig2} shows that the density fluctuations of matter are more strongly affected by a PMF for wavenumbers $k/h >  0.1$ Mpc$^{-1}$\cite{Yamazaki:2006mi}. 
  This is because the energy density of the PMF, $E_\mathrm{B}$, only depends  upon the scale factor as $a^{-4}$ like the photons, and unlike the photons, the magnetic field fluctuations can survive on scales below the photon diffusion length, i.e.~below the Silk damping scale. 
Since the CDM and baryons interact with each other through gravity, 
the Lorenz force from the PMF can also indirectly affect the cold dark matter (CDM). This effect on the CDM is much smaller than that on the baryons before the epoch of photon last scattering because the density of baryons oscillates with that of photons and their gravitational effect on the CDM is very small.
After the epoch of photon last scattering, the time evolution of the CDM density starts to be influenced by the baryon density through gravitational interaction \cite{Yamamoto:1997qc}. 
Therefore, the Lorenz force effect of the PMF on the CDM increases with time.

The strength of this effect is dependent upon the ratio of the baryon density to the CDM density, $\Omega_b/\Omega_c$.
The density fluctuations of the baryons  are directly generated by the PMF, while  the density fluctuations of the CDM are indirectly generated from the gravitational interaction with the baryons.  Therefore,  the density fluctuations of the CDM only grow due to  gravitational instability, and their growth rate is nearly the same as the density fluctuations of the primordial CDM. 
On the other hand, 
the evolution of the baryon density fluctuations is not dominated by
the gravitational potential 
until the gravity becomes comparable to the Lorentz force, i.e. 
$4\pi G \rho \, \delta \sim k^2 \frac{\Pi_{\mathrm{[EM:S]}}(\mbi{k},\tau)}{4\pi \rho_b}$, where $\Pi_{\mathrm{[EM:S]}}(\mbi{k},\tau)$ is the Lorentz force power spectrum \cite{Yamazaki:2006mi}.  This is because the baryons density fluctuations are generated by the PMF directly.
As mentioned above,  the effect of a PMF on the baryon density fluctuations  is
largest on the  smallest scales (larger $k$ values).  Hence, 
 the density fluctuations for  baryons in the presence of a  PMF increase with the
wavenumber $k$\cite{Yamazaki:2006mi}.
\subsection{Matter Power Spectra with a PMF }
It should be noted that the effect of a PMF on the matter power spectrum, $P(k)$ is different from effect of a PMF on the matter density fluctuations $\delta$. 
The fluctuations of total density $\delta$ can become smaller or larger depending upon whether the dominant effect is from the pressure or the tension of the PMF. 
On the other hand, the effects of a  PMF upon the matter power spectrum are  dependent on how well the spectrum of the PMF correlates with the primary density fluctuations. 

Figure \ref{fig2} illustrates the degeneracy between the PMF-matter-density correlation
 factor $s_\mathrm{[DF]}$  
and the PMF spectral index $n_B$.
This figure shows that a change of spectral index from $n_B = -2.0$ to $-1.0$ can be offset by including a negative correlation.
The correlation factor $s_\mathrm{[DF]}$, however,  depends upon the origin of the PMF.
We will discuss the  below the resolution of this degeneracy problem, along with the relation between  the correlation factor $s_\mathrm{[DF]}$ and
the origin of the PMF. 
\subsection{\label{ssec:ne}Negative correlation of the density perturbations}
In the case of a negative correlation between
the PMF and the matter density fluctuations,
the pressure of the PMF accelerates the evolution of the fluctuations in the  matter density, while 
the tension of the PMF delays them.
Thus, for $n < -1.5$, for which the pressure of the PMF dominates,
the PMF accelerates the evolution of the matter density fluctuations. 
On the other hand, for $n > -1.5$, for which the tension dominates, the PMF delays the evolution.
If we assume that the PMF would have been produced in  low density regions, such negative correlations might be allowed.
This situation, however, is difficult to realize for a  causally produced PMF,  because it seems  natural that more amplitude for the PMF would be generated in regions with 
higher matter density and therefore higher currents.
\subsection{\label{ssec:no}No correlation}
The PMF and the matter density fluctuations are uncorrelated when $s_\mathrm{[DF]}=0$. 
The PMF then increases the matter power spectrum independently of whether the  pressure or tension dominate the PMF.
If density fluctuations  generate the PMF as in Ref.~\cite{ichiki:2006sc}, 
peaks in the  amplitude of the PMF should  lie at peaks in the pressure
gradient of  the cosmological fluids. 
In this case, the PMF is produced along the
border between high and low density regions, i.e., $\delta\sim0$.
Eventually, such a PMF has no (or very weak) statistical correlation with fluctuations in the  matter density.
\subsection{\label{ssec:p}Positive correlation}
In the case of a positive correlation between
the PMF and  fluctuations in the  matter density,
the evolution of the matter density fluctuations is accelerated by the tension of the PMF.  Therefore,
 the matter evolution is delayed by the pressure of the PMF.
Thus, for $n < -1.5$, for which the pressure of the PMF dominates,
the evolution of the matter density fluctuations is delayed by the PMF. 
On the other hand, for $n > -1.5$, a PMF accelerates the evolution.
If we assume that the PMF would have been produced in  higher density regions, such positive correlations might be allowed.
\section{Effects of a PMF on the CMB}
In this section, we review effects of a PMF on the temperature fluctuations and polarization anisotropies of the CMB (see \cite{Yamazaki:2007oc,Yamazaki:2009na} for details). 
Regarding the CMB polarization, the reader should be aware that in cosmology the intensity of polarized radiation is expressed in terms of two scalar fields $E$ and $B$ that are independent of how the coordinate system is oriented. E and B are the curl-free and curl-like components of the linear polarization field. Based upon these there are three types of fluctuations: scalar, vector and tensor.  There are then four observables: the temperature, $E$-mode, $B$-mode, and the temperature cross polarization power spectra.
From these, one can generate three power spectra TT, EE, and BB, where T is the total intensity  and also three cross-spectra TE, TB, and EB. However, the only nonzero spectra are  TT, EE, BB, and TE due to parity considerations. 

\begin{figure*}[h]
\includegraphics[width=1.0\textwidth]{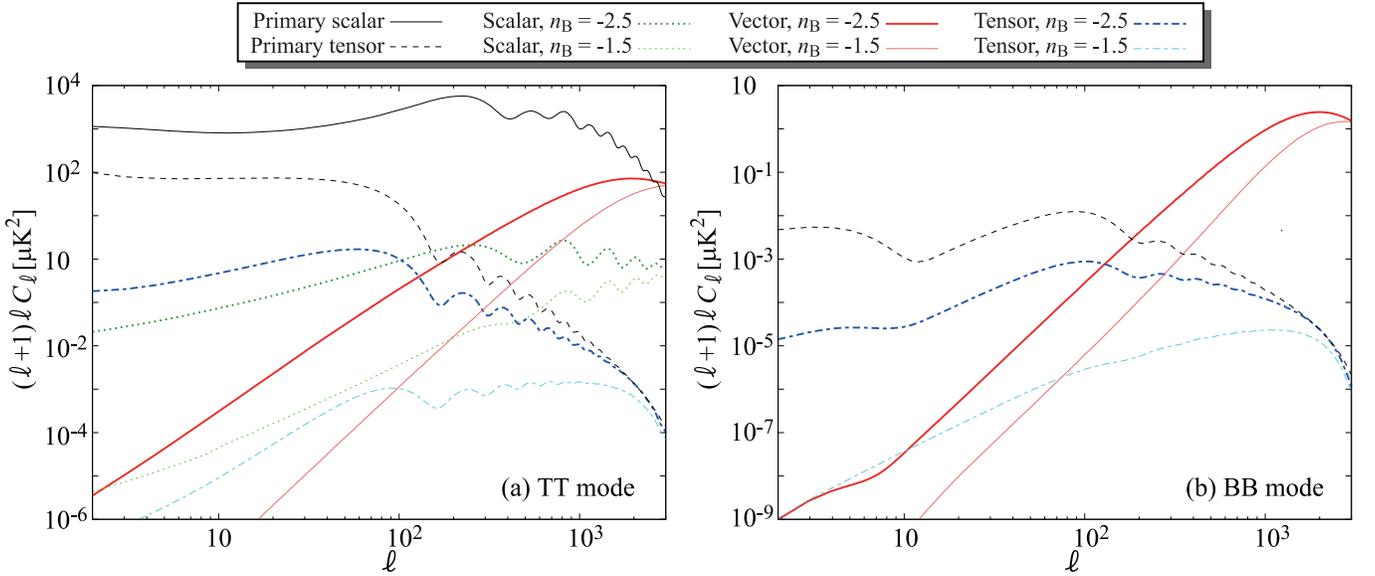}
\caption{\label{figTTBB} 
Polarization anisotropies and temperature fluctuations of CMB the from a PMF. 
Panels (a) and (b) show TT and BB modes, respectively, for models with $B_\lambda = 3.0$ nG and $n_\mathrm{B} = -1.5$ or -2.5 as labeled. 
We use the scalar to tensor ratio of 0.2 for the primary tensor.
}
\end{figure*}

\begin{figure*}[h]
\includegraphics[width=1.0\textwidth]{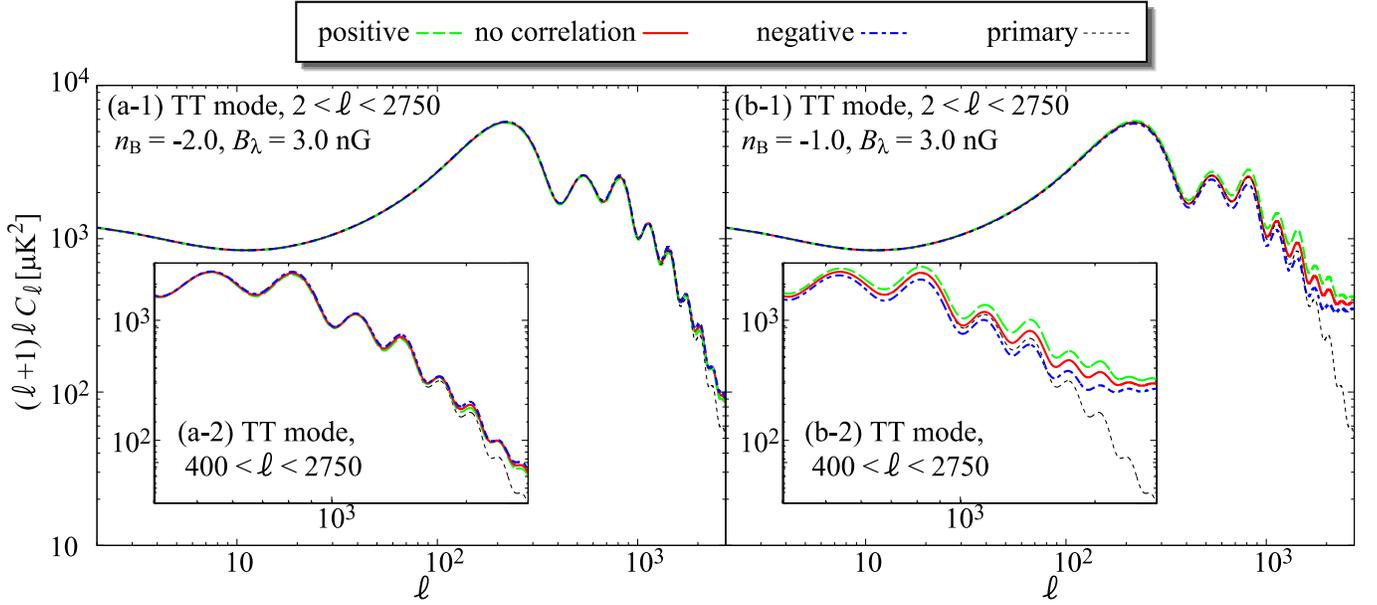}
\caption{\label{fig3} 
CMB temperature fluctuations from the PMF. 
Plots show various ranges for: (a-1) and (b-1) TT($2 < \ell < 2750$), and (a-2) and (b-2) TT($400 < \ell < 2750$) with $B_\lambda = 3.0$ nG and $n_\mathrm{B} = -2.0$ (left panel) or $-1.0$ (right panel). 
Green dashed, red bold, blue dashed-dotted, and black dotted curves show positive, no and negative correlations, respectively. The black dotted curve shows the  primary spectrum without a PMF.
}
\end{figure*}

\begin{figure*}[h]
\includegraphics[width=1.0\textwidth]{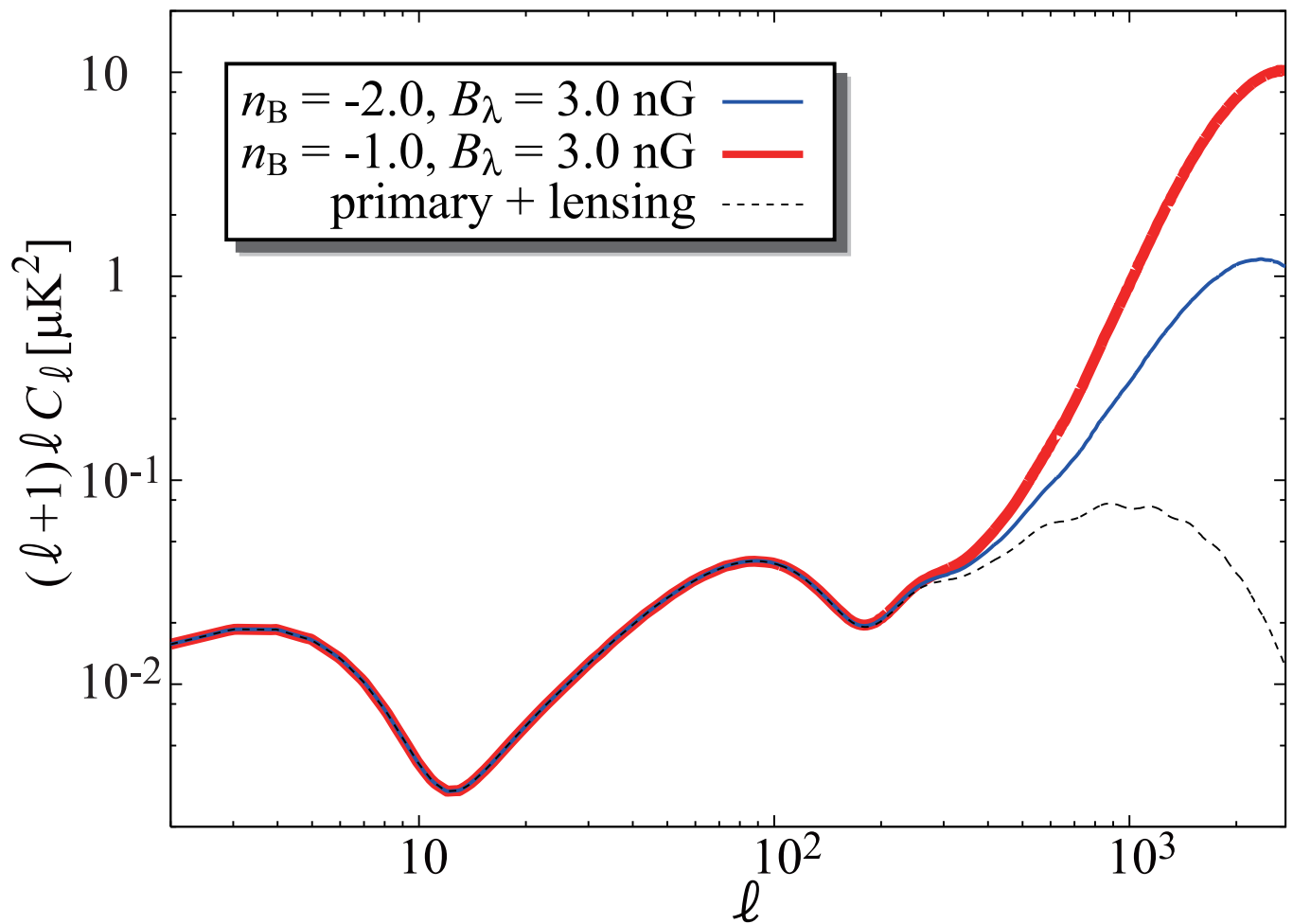}
\caption{\label{fig4} 
CMB polarization anisotropies from the PMF.
Blue thin and red bold curves show $(B_\lambda, n_\mathrm{B}) = (3.0 \mathrm{nG}, -2.0)$ and $(B_\lambda, n_\mathrm{B}) = (3.0 \mathrm{nG}, -1.0)$. The black dotted curve shows a primary spectrum with weak lensing effects (without the PMF).
}
\end{figure*}

Figures \ref{figTTBB} - \ref{fig4} show that the usual (TT mode) CMB power spectra for large multipoles $\ell$ (small angular scale) are influenced most strongly by the PMF. 
The first reason for this  is that the energy density of the PMF   scales with the cosmic expansion as   $a^{-4}$ just like the photons,  Unlike  photons, however, the magnetic fields are not much affected by photon diffusion. As a result, the temperature fluctuations and polarization anisotropies of the CMB continue to be affected by a PMF even for angular scales smaller than  the Silk damping scale.
Another  reason is that a dominant contribution to  the CMB on smaller angular scales comes from the vector mode of the PMF.
This is because, after horizon crossing, the scalar mode cannot evolve due to
the acoustic oscillations.  That is, the photons fall in and out of gravitational potentials but cannot grow. On the other hand, the vector mode of the PMF fluctuations can continue growing once inside the horizon.
For higher $\ell$, the vector mode also dominates the CMB polarization from the PMF, and the tensor mode from the PMF diminishes [cf.figures \ref{figTTBB} - \ref{fig4}]\cite{Seshadri:2000ky,Lewis:2004ef,Yamazaki:2007oc,Yamazaki:2009na}.
The gravitational waves generated by the PMF can be negligibly small after horizon crossing.  This is  because this homogeneous solution starts to oscillate inside the horizon and decay rapidly
\cite{
1979ZhPmR..30..719S,
1982PhLB..115..189R,
1985SvA....29..607P,
Pritchard:2004qp}. 
Consequently, the anisotropy spectrum is affected by gravity waves only on
scales larger than the horizon at the epoch of photon last scattering.  

Panel (b) on figure \ref{figTTBB} and figure \ref{fig4} show that the BB mode from the PMF can dominate for $B_{\lambda}\gtrsim 2.0$ nG for multipoles with  $\ell \gtrsim 200$.  On these angular scales  there is also the BB mode which is converted from the  EE power spectrum by  gravitational lensing  \cite{Pritchard:2004qp}. Therefore, there is a degeneracy between the PMF and gravitational lensing on small angular scales.
However, since the spectrum from the effects of the gravitational lensing signal have been independently determined, this power can be subtracted directly.
 We have shown \cite{Yamazaki:2007oc, Yamazaki:2009na} that the effects of a PMF on the temperature fluctuations of the CMB are stronger with correlations  than than without,  as is also  the casefor the matter power spectrum with a PMF. This is illustrated in Figure  \ref{fig3}.

There is also a degeneracy between the Sunyaev-Zel'dovich effect and the PMF.
Nevertheless, the effects of a PMF on the CMB are independent of frequency because
the PMF influences the primary CMB as a background. 
Eventually, by using observations at different frequencies, it should be possible to distinguish the effects of a PMF  from foreground effects which depend upon frequency.
\section{Constraints on the PMF}
In this section, we  review  constraints on the parameters of the PMF from cosmological observations. First, we  show that a strong  constraint on the PMF amplitude and spectral index can be  derived from the prior observed constraints on  the $\sigma_8$ parameter.  This fiducial quantity represents  the root-mean-square of the matter density fluctuation in a comoving sphere of radius $8h^{-1}$ Mpc and as such is a measure of the growth of large scale structure. It is given by a weighted integral over  the matter power spectrum \cite{Peebles:1980booka}. Second, we note  that the parameters of the PMF can be strongly constrained by using the observational data on the CMB with the previous priors used in the analysis of Refs.~\cite{Spergel:2006hy,Hinshaw:2006ia,Page:2006hz} without a PMF. In the last subsection, we will discuss the possible origins of a PMF based upon the deduced constraints on the parameters characterizing the PMF.
\subsection{Prior on the PMF Parameters from $\sigma_8$}
Because of the degeneracy of  affects on the observed CMB power spectrum from various parameters, it has become standard in the field of CMB physics to fit the CMB parameters by a Baysian method that makes optimum use of the varity of independently determined constraints on cosmological parameters.  For this purpose the Monte-Carlo Markov Chain (MCMC)\cite{Lewis:2002ah} analysis has become the method of choice to find the multidimensional surface of maximum likelihood among the many parameters.  In this context we note that the PMF amplitude $B_\lambda$ and power law index $n_\mathrm{B}$ have a strong degeneracy.  Therefore, one must identify a strong prior constraint to effectively decouple these two  parameters of the PMF.  
In this quest we are aided by observed constraints on the  $\sigma_8$ parameter for which a value of $0.7 < \sigma_8 < 0.9$ has been deduced from diverse observational data on linear cosmological scales\cite{Cole:2005sx,Tegmark:2006az,Rozo:2007yt,Ross:2008ze}.  Using this range as a prior in the Monte-Carlo Markov Chain (MCMC)\cite{Lewis:2002ah} analysis,   constraints on the parameters of the PMF can be obtained by  a fit  to the observed power spectra of the CMB.
For some of the cosmological parameters, $\Omega_b$, $\Omega_\mathrm{CDM}$, $n_\mathrm{S}$, and $A_\mathrm{S}$ affect $\sigma_8$, we also have to consider the possible degeneracy between these cosmological parameters and the PMF parameters.
Fortunately, however, recent observations of the CMB have constrained such cosmological parameters on larger scales ($\ell < 1000$) \cite{Spergel:2006hy,Hinshaw:2006ia,Page:2006hz}.  As mentioned above,  the PMF mainly affects the CMB anisotropies on smaller scales ( $\ell > 1000$)\cite{Yamazaki:2006bq,Yamazaki:2006mi,Yamazaki:2007oc}.
Therefore, only a  small degeneracy between the PMF parameters and the other cosmological parameters is expected, so that fixing the other cosmological parameters at the WMAP best fit values is justified. 

\begin{figure*}[h]
\includegraphics[width=1.0\textwidth]{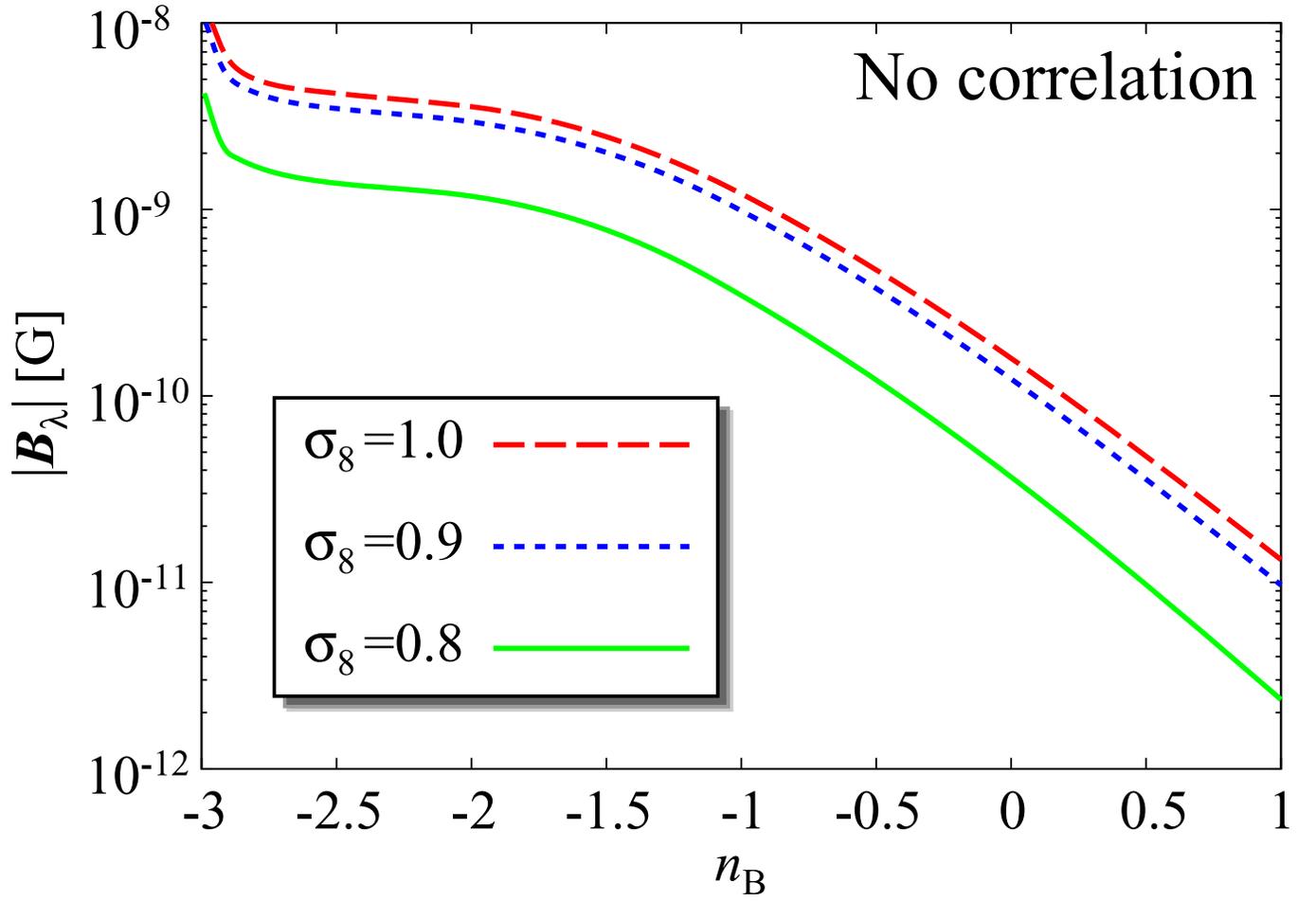}
\caption{\label{fig5} 
Curves of constant values for the $\sigma_8$ in the parameter plane of PMF amplitude $B_\lambda$ vs. spectral index $n_\mathrm{B}$ for the case of no correlation between the PMF and the matter density fluctuations. Red dashed, blue dotted and green bold curves show constant values of $\sigma_8 =1.0, 0.9$ and 0.8, respectively.
}
\end{figure*}

\begin{figure*}[h]
\includegraphics[width=1.0\textwidth]{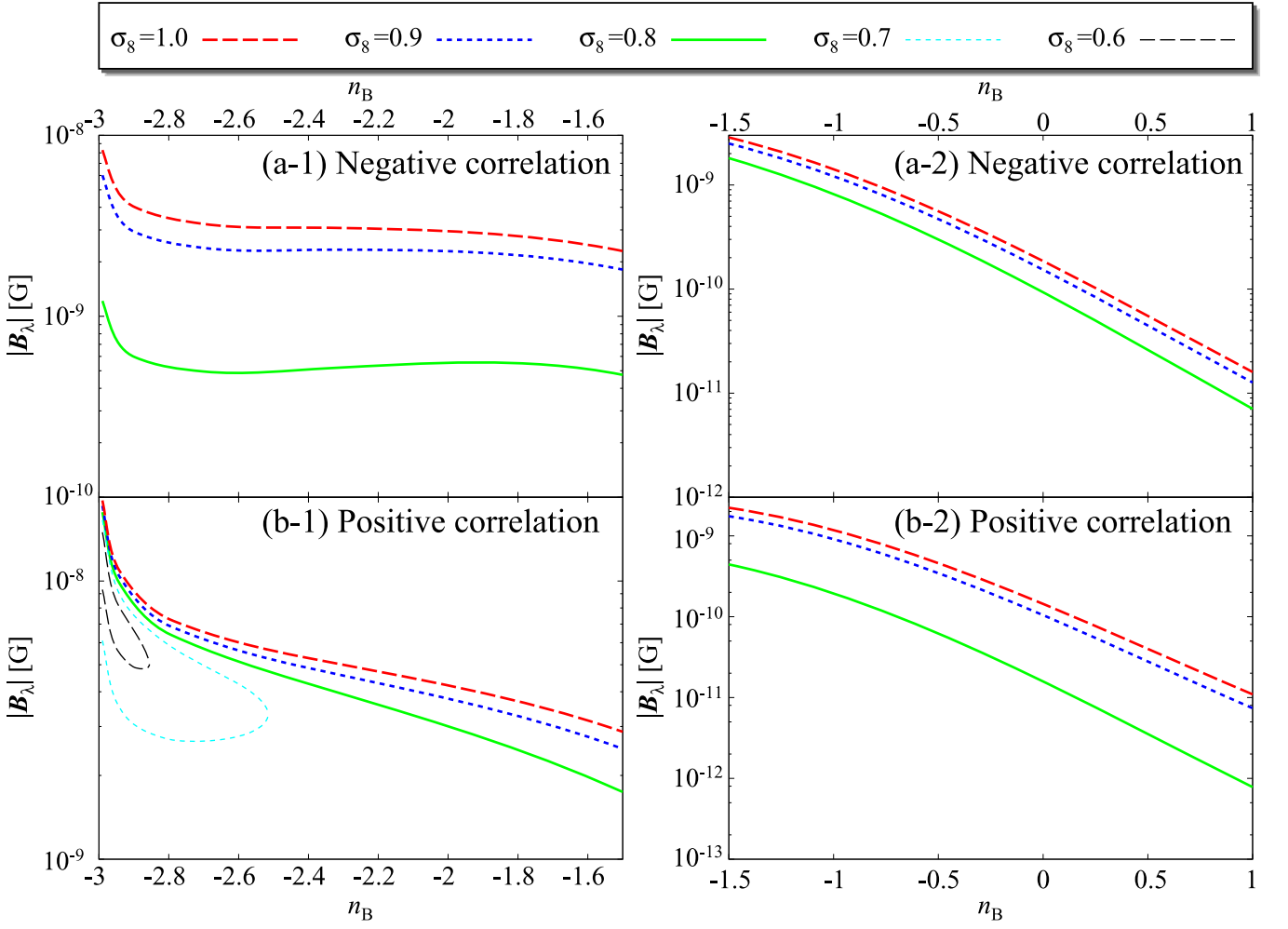}
\caption{\label{fig6} 
Curves of constant values for the $\sigma_8$ in the parameter plane of PMF amplitude $B_\lambda$ vs. spectral index $n_\mathrm{B}$ for the negative($s_\mathrm{[DF]}=-1$) and positive($s_\mathrm{[DF]}=1$) correlations between the PMF and the matter density fluctuations. 
Panels (a-1) and (b-1) show $n_\mathrm{B} < -1.5$ as the PMF pressure dominated region.
On the other hand, panels (a-2) and (b-2) show $n_\mathrm{B} > -1.5$ as the PMF tension dominated region.
Red dashed, blue dotted, green bold, azure dotted thin, and black dashed thin curves show constant values of $\sigma_8 =1.0, 0.9, 0.8, 0.7$ and 0.6, respectively.
}
\end{figure*}

Figure \ref{fig5} and the panels in Figure \ref{fig6} show the optimum PMF parameters $n_\mathrm{B}$ and $B_\lambda$ for various constant values of $\sigma_8$ as labeled.
Since the PMF power spectrum depends upon $k^{2n_\mathrm{B}+3}$\cite{Yamazaki:2007oc},  for $n_\mathrm{B} < -1.5$, the PMF effects on the density fluctuations for smaller scales decrease with lower values for $n_\mathrm{B}$.
Therefore since $\sigma_8$ depends upon the amplitude of the matter power spectrum for smaller scale, when the spectral index is near $n_\mathrm{B} = -3.0$, the matter power spectrum in the presence of a  PMF is smaller for smaller scales. Thus, larger amplitudes of $B_\lambda$ are allowed.
For $n_\mathrm{B} > -1.5$, however,  the energy density power spectrum of the PMF is proportional to  the cut-off scale $k_\mathrm{C}^{2n_\mathrm{B}+3}$\cite{Yamazaki:2007oc}. 
$k_\mathrm{C}$  is  also, proportional to $B_\lambda^{-1/(n_\mathrm{B}+5)}$ \cite{Jedamzik:1996wp,Subramanian:1997gi,Banerjee:2004df,Yamazaki:2007oc}.
Substituting these relations into Eqs.(15)-(17) in Ref. \cite{Yamazaki:2007oc}, we easily obtain the following relation between the PMF power spectrum and the magnetic field strength,  $P(k)_\mathrm{PMF} \propto B_\lambda^{{[14}/{(n_\mathrm{B}+5})]}$.
Considering $B_\lambda \ll 1$G, the matter power spectrum with the PMF for $n_\mathrm{B} > -1.5$ becomes larger for larger $n_\mathrm{B}$
Therefore larger amplitudes of $B_\lambda$ are not allowed for larger $n_\mathrm{B}$.

Since the effects of a PMF for negative and positive correlations change at $n_\mathrm{B} = -1.5$ (see Section \ref{s:transfer_function}), we divide the discussion below of each correlation into two parts based upon whether the spectral index is greater or less than  $n_\mathrm{B} = -1.5$.
\subsubsection{No Correlation}
As mentioned above in Sub-Section.\ref{ssec:no}, 
in the case of no correlation between the PMF and the density fluctuations from primary perturbations, the effect of a PMF is to increase the matter power spectrum independently of whether the pressure or tension dominates the PMF.
As noted above, from recent constraints on the $\sigma_8$ parameter from cosmological observations of LSS  we can exclude PMF parameters which imply $\sigma_8 > 1$.

Hence, we can exclude a PMF with field strength  $B_\lambda\ ^>_\sim 1$ nG if $n_\mathrm{B} > -0.9$ (Fig.~\ref{fig5}).  We can also exclude a  PMF with amplitude $B_\lambda\ ^>_\sim 0.1$ nG if $n_\mathrm{B} > 0.2$. 
If we assume that the origin of the observed magnetic field in the clusters of galaxies is the PMF, we can limit the PMF amplitude to $B_\lambda\ ^>_\sim 1$ nG. In this case, we can exclude a PMF spectral index in the range $n_\mathrm{B} > -0.9$.
\subsubsection{Negative Correlation\label{n_c}}
As noted above in subsection \ref{ssec:ne}, in the case of a negative correlation,
the tension of the PMF delays the evolution of the matter density fluctuations, while 
 the pressure  tends to accelerate them. Furthermore, the pressure of the PMF dominates for $n_\mathrm{B} < -1.5$, while the tension of the PMF dominates for $n_\mathrm{B} > -1.5$.
Therefore, for $n < -1.5$ the matter density fluctuations are accelerated by the PMF, while, for $n > -1.5$, this evolution is delayed by the PMF.
This behavior can be traced to the third terms in both Eqs.(\ref{crossb}) and (\ref{crossCDM}).
Using the allowed range of PMF parameters as mentioned above,
we can exclude the range of PMF amplitude to $B_\lambda\ ^>_\sim 1$ nG if $n_\mathrm{B} > -0.81$, and the range of PMF amplitudes to $B_\lambda\ ^>_\sim 0.1$ nG if $n_\mathrm{B} > 0.26$(panels (a-1) and (a-2) in Figure \ref{fig6}). 
However, as mentioned above, it is difficult to model the production of  a PMF in matter fields of low energy density in order to realize such negative correlations. 
\subsubsection{Positive Correlation}
As noted above in Sub-Section.\ref{ssec:p}, in the case of a negative correlation,
the pressure of the PMF delays the evolution of the matter density fluctuations
while the tension  accelerates them. Considering the  same conditions as in the previous section.V A 1, the PMF decreases the fluctuations of matter density for $n_\mathrm{B}< -1.5$. On the other hand, for $n_\mathrm{B}> -1.5$, it increases the fluctuations. In this case, we can exclude the range of PMF amplitudes with $B_\lambda\ ^>_\sim 1$ nG if $n_\mathrm{B} > -0.94$, and the range of PMF amplitudes with $B_\lambda\ ^>_\sim 0.1$ nG for  $n_\mathrm{B} > 0.13$ (panels (b-1) and (b-2) in Figure \ref{fig6}). 
\subsection{Constraint on the PMF from the CMB}
In this section, we summarize  the current constraints on parameters of the PMF from fits to  the CMB  and LSS observational data.
We consider a flat ($k = 0$) $\Lambda$CDM universe characterized by 8 parameters, i.e.~$\{ \Omega_b h^2,
	\Omega_c h^2,
	\tau_C,
	n_s,$
	$\log(10^{10}A_s),
	A_t/A_s$, 
	$|B_\lambda|$,
	$n_\mathrm{B} \}$, 
where
$\Omega_b h^2$
and
$\Omega_c h^2$
are the baryon and CDM densities in units of the critical density,
$h$ denotes the Hubble parameter in units of 100 km s$^{-1}$Mpc$^{-1}$,
$\tau_C$
is the optical depth for Compton scattering,
$n_s$
is the spectral index of the primordial scalar fluctuations,
$A_s$
is the scalar amplitude of the primordial scalar fluctuations and
$A_t$
is the scalar amplitude of the primordial tensor fluctuations.
We define the tensor index for the primordial tensor fluctuations as $n_t =-(A_s/A_t)/8 $.
For all cosmological parameters we use the same prior constraints as those adopted in the WMAP analysis \cite{Dunkley:2008ie}.
As noted in the previous section, the case of no correlation is more realistic than  the case of negative or  positive correlations. Therefore, we focus here on the constraints of the parameters characterizing  the PMF in the case of no correlation.

Using the MCMC algorithm and available cosmological observations, we have constrained the standard cosmological parameters and the PMF parameters listed in Table.1.   We note, however, that the MCMC algorithm tends to sparsely sample near a boundary, and hence find  low probability there.  This is why it appears that $n_B$ is constrained even  in the limit that $B_\lambda \rightarrow 0$.  Of course, for $B_\lambda  = 0$ there is no constraint on the spectral index at all.  This shortcoming, however,  does not negate the fact that there is a genuine minimum in the goodness of fit for a finite magnetic field and spectral index.  In the optimum fit with the inclusion of parameters of the PMF, the minimum total $\chi^2$ changes from 2803.4 to 2800.2 corresponding to a change in the $\chi^2$ per degree of freedom from 1.033 to 1.031. Therefore, a finite  PMF slightly improves the goodness of fit even after taking into account the new degrees of freedom.  The existence of a PMF, however, is still only of marginal significance.

\begin{figure*}[h]
\includegraphics[width=0.9\textwidth]{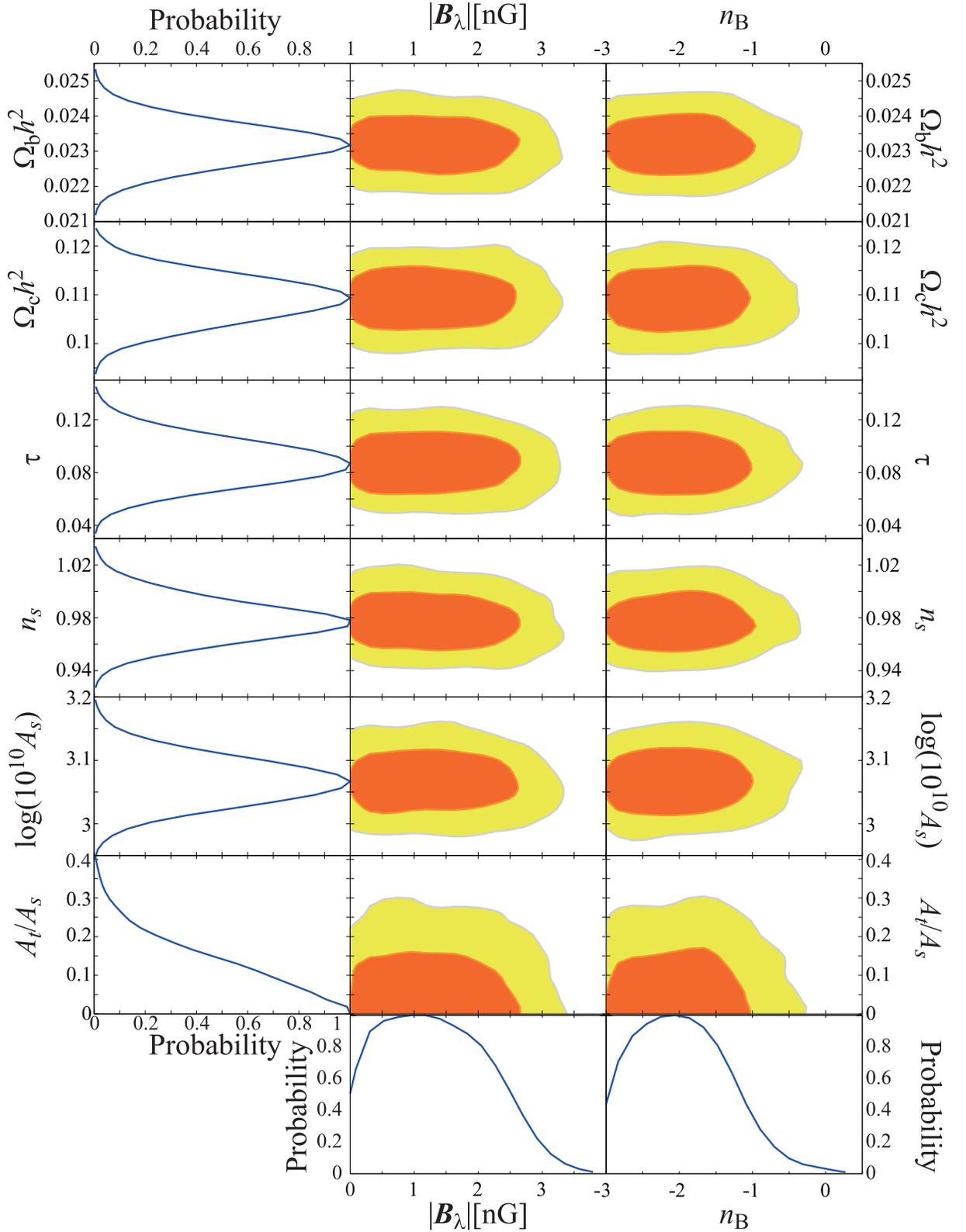}
\caption{\label{fig7}
Probability distributions and contours of 1$\sigma$ and 2$\sigma$ confidence limits for the standard cosmological parameters as a function of the PMF field strength $|B_\lambda|$ and power law index $n_\mathrm{B}$.  
Orange contours show 1 $\sigma$(68\%)
 confidence limits and yellow contours show 2 $\sigma$(95\%)
confidence limits. 
Blue curves in the left and bottom of the figure show the probability distributions of each parameter.  Note the existence of a maximum in the probability distributions for $|B_\lambda|$  and $n_\mathrm{B}$. 
The standard cosmological parameters do not have a degeneracy with the PMF parameters because they are mainly constrained by the observed CMB data for $\ell<$ 1000 (up to  the 2nd peak), while the PMF is mainly influenced by the power on smaller angular scales and higher multipoles, $\ell>$ 1000.
}
\end{figure*}

\begin{figure*}[h]
\includegraphics[width=1.0\textwidth]{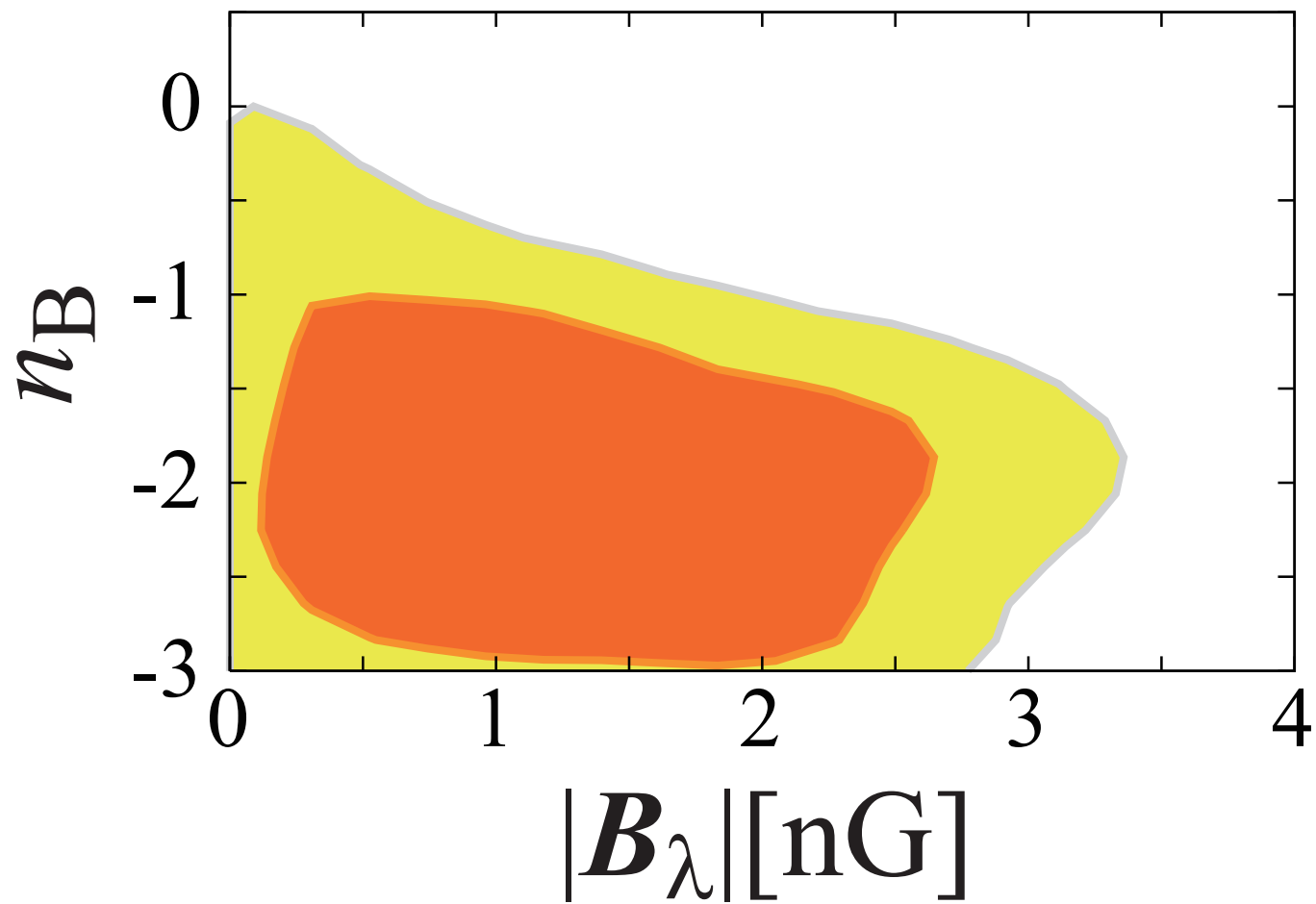}
\caption{\label{fig8} Probability contours in the plane of $B_\lambda$ vs. $n_\mathrm{B}$. Orange contours show 1 $\sigma$(68\%)
 confidence limits and yellow contours show 2 $\sigma$(95\%)
confidence limits. 
}
\end{figure*}

Figures \ref{fig7} and \ref{fig8} show the 68\% and 95\% C.L.~probability contours in the planes of
 various cosmological parameters versus the amplitude $|B_\lambda|$, or power law index $n_\mathrm{B}$, along with the probability distributions. 
Figure.~\ref{fig8} and the bottom panels in Fig.~\ref{fig7} show the probability distributions for $|B_\lambda|$  and $n_\mathrm{B}$.  
Of particular note for this review is the presence of maxima in the likelihood functions for $|B_\lambda| = 0.85 \pm 1.25$ nG and  $n_\mathrm{B}= -2.37^{+0.88}_{-0.73}$. 
Although these values are consistent with zero magnetic field, and  thus only imply upper limits, they suggest the possibility that with forthcoming data (particularly for large CMB multipoles) a finite magnetic field may soon be detectable.

These figures exhibit no degeneracy between the PMF parameters and the standard  cosmological parameters. Table 1 confirms that the standard cosmological parameters are not significantly different from those deduced directly from the WMAP 5yr data analysis without a PMF.
The reason for this is simple. The standard cosmological parameters are mainly constrained by the observed CMB power spectrum for low multipoles $\ell < 1000$ (up to the 2nd acoustic peak).  On the other hand, the PMF dominates for 
$\ell > 1000$.   Hence, the  PMF effect on the power spectrum is  nearly independent of the standard cosmological parameters.

The tensor to scalar ratio $A_t/A_s$ deduced in our analysis is smaller than the upper limit $A_t/A_s <$ 0.43 (95\% CL) deduced from the WMAP 5yr data analysis without a PMF. 
The reason for this is that we define $A_t$ as the tensor amplitude of the primary CMB spectrum (without a PMF).  This tensor term only arises from the primordial gravitational-wave background produced during inflation. We combine the tensor amplitude from the PMF and $A_t$ when we compare our tensor amplitude with the result deduced by others. The value of $A_t$ by itself is comparable  to the tensor contribution from the  PMF.  The value of $A_t/A_s$ from the WMAP 5yr data is less than 0.43 (95\% CL).  Hence, 
our result  is consistent with the previous constraints when the additional PMF contribution is included.

The degeneracy of the PMF parameters \cite{Yamazaki:2006bq} is broken by the different effects of the PMF on both the matter power spectrum and the CMB power spectrum.
The vector mode can dominate for higher $\ell$ in the CMB temperature anisotropies \cite{Yamazaki:2007oc}, while the matter power spectrum becomes sensitive \cite{Yamazaki:2006mi} to the power law spectral index $n_\mathrm{B}$ when a PMF is present.
Additionally, the CMB fluctuations from the PMF are smaller than the primary CMB fluctuations for the scalar and tensor modes on large angular scales \cite{Paoletti:2008ck,Finelli:2008xh}. 
Therefore, the tensor to scalar ratio is not affected by the PMF.

\begin{figure*}[h]
\includegraphics[width=1.0\textwidth]{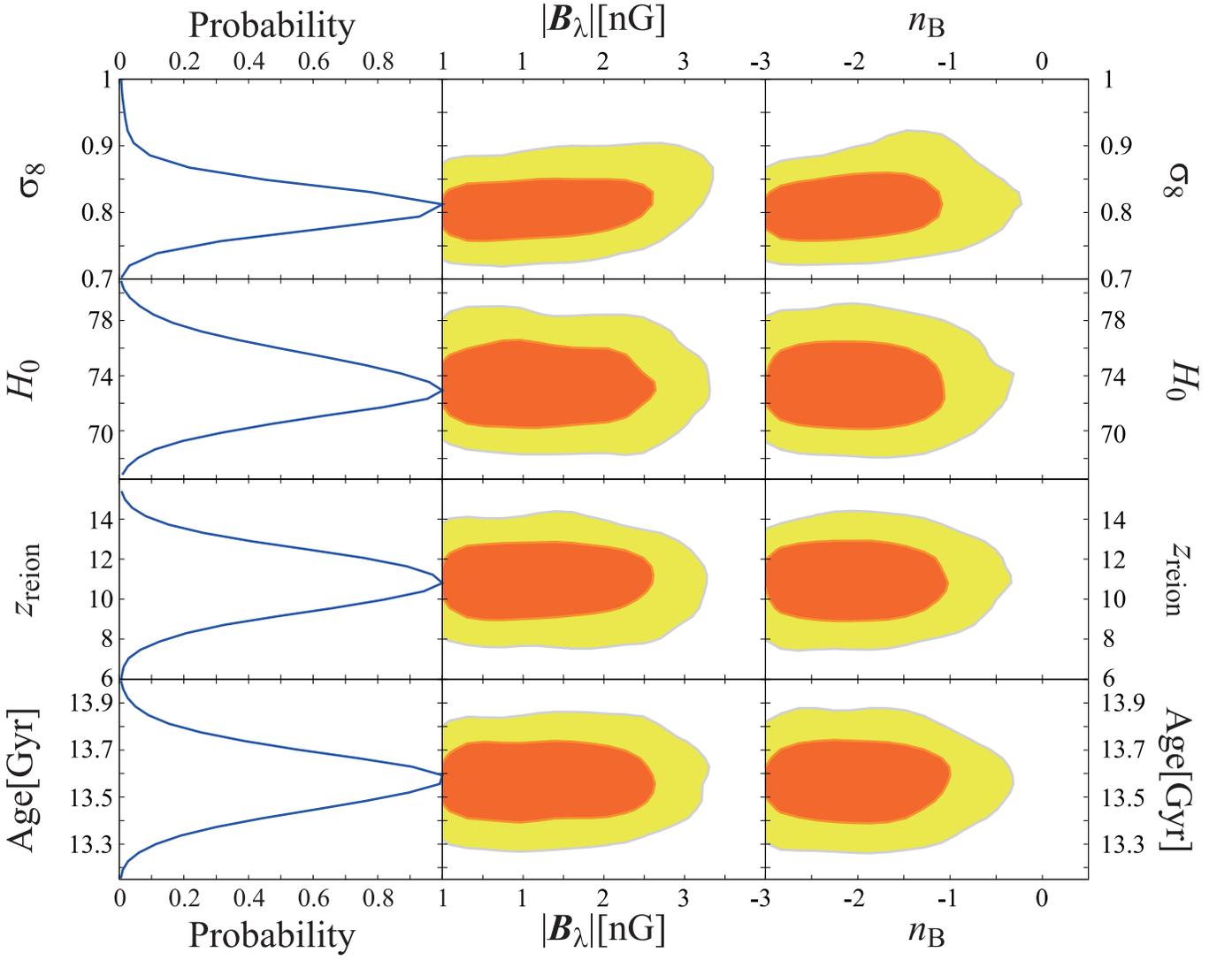}
\caption{\label{fig9} 
Probability distributions and contours of 1$\sigma$ and 2$\sigma$ confidence limits for the parameters  $\sigma_8$, $H_0$, $z_\mathrm{reion}$ and Age. 
Orange contours show 1 $\sigma$(68\%)
 confidence limits and yellow contours show 2 $\sigma$(95\%)
confidence limits. 
Blue curves show probability distributions for each parameter.
Note that these are not input priors,  but output results.
}
\end{figure*}

Figure \ref{fig9} shows our deduced probability distributions and the 1$\sigma$ and $2\sigma$ (68\% and 95\% C.L.)
probability contours for the resultant cosmological parameters, $\sigma_8$, $H_0$, $z_\mathrm{reion}$, and Age.  Here, 
$H_0$ = 100$h$ km s$^{-1}$ Mpc$^{-1}$ is the Hubble parameter, $z_\mathrm{reion}$ is the red shift at which re-ionization occurs,  and Age is the presently observed age of the universe in Gyr.
It is important to keep in mind that these parameters are not  input parameters, but are  output results.  The sum of the results on Figures \ref{fig7} - \ref{fig9} provides marginal evidence that both upper and lower limits to the parameters of the PMF can be deduced.

Table 1 summarizes the upper limits to the PMF parameters with input and output cosmological parameters, $\sigma_8$, $H_0$, $z_\mathrm{reion}$, and Age.  
In particular we find that 
$|B_\lambda| \mathbf{< 2.10} ( 68\% \mathrm{CL})$
nG and $\mathbf{< 2.98} ( 95\% \mathrm{CL})$ nG and  
$n_\mathrm{B} \mathbf{< -1.19} ( 68\% \mathrm{CL})$
and 
$\mathbf{< -0.25} ( 95\% \mathrm{CL})$ at a present scale of 1 Mpc.
Although previous work  
\cite{Yamazaki:2006bq,Yamazaki:2007oc,Yamazaki:2006mi,Yamazaki:2008bb}  could obtain a less stringent  upper limit to $|B_\lambda|$ they could not constrain $n_\mathrm{B}$ at all. Moreover, our deduced probability distributions  suggests that a finite PMF provides the best fit.  

On angular scales smaller than that probed by the CMB the observed number density of galaxies is a better measure of the power spectrum.  Therefore, since the PMF mainly influences the  small angular scales, using the combined LSS observational data  (2dFDR\cite{Cole:2005sx}), and the CMB
(WMAP 5yr\cite{Hinshaw:2008kr},
ACBAR\cite{Kuo:2006ya},
CBI\cite{Sievers:2005gj},
Boomerang\cite{Jones:2005yb})  
 we can  constrain the PMF better than in previous works which relied on the CMB data only. 
In particular, an upper limit on $n_\mathrm{B}$ can be constrained for the first time while the lower limit to $n_\mathrm{B}$ is approaching the  1$\sigma$ confidence level. 

The right-bottom panel of Figure \ref{fig7} shows that the maximum likelihood is for a spectral index of $n_\mathrm{B}= -2.4^{+0.9}_{-0.7}$.
It is important\cite{Caprini:2001nb,Yamazaki:2006bq} to constrain  $n_\mathrm{B}$ as this parameter provides insight into models for the formation of the PMF.   
If the PMF were formed during inflation one would expect a scale-invariant value of  
 $n_\mathrm{B} = -3$.  The value deduced in Figures \ref{fig6}-\ref{fig9} and Table 1, are thus consistent with an inflation generated PMF at the 1$\sigma$ confidence level.  
 
Next we discuss other generation models for the PMF during  various phase transitions 
\cite{Vachaspati:1991nm,Kibble:1995aa,Ahonen:1997wh,Joyce:1997uy} 
from the gravity waves produced along with the PMF. The formation of light elements during big-bang nucleosynthesis (BBN) depends upon a balance between the nuclear reaction rates and the expansion rate of the universe. Since gravity waves contribute to the total energy density they affect the expansion rate. Hence, they are constrained by a comparison between the BBN predictions and the observed light element abundances \cite{Caprini:2001nb,Yamazaki:2006bq}.
There are also other  theoretical constraints on the origin of the PMF. For example, in Ref.~\cite{Durrer:2003ja}  it was noted  that  a
non-scale-invariant spectrum of the PMF is difficult to generate or survive if it is formed via a  causal mechanism  during later phase transitions.  Indeed, as pointed out in Ref.~\cite{Caprini:2001nb,Durrer:2003ja} the spectral  index of a magnetic field generated during any period of standard Friedmann  expansion (i.e. any time except inflation) is constrained to be a positive even integer larger than two: $n_B=2, 4, 6 ....$. This  constraint, which applies for example to a PMF generated at a phase transition, or by  second order MHD processes and so on, comes simply from the fact that the magnetic field generation is a causal process (i.e. the universe has a finite horizon at the magnetic field generation time).   Based upon this, the fact our results favor a negative spectral index near -3, indicate that an inflation mechanism \cite{Turner:1987bw,Ratra:1991bn,Bamba:2004cu} may be the most likely origin for the  PMF.

\begin{figure*}[h]
\includegraphics[width=1.0\textwidth]{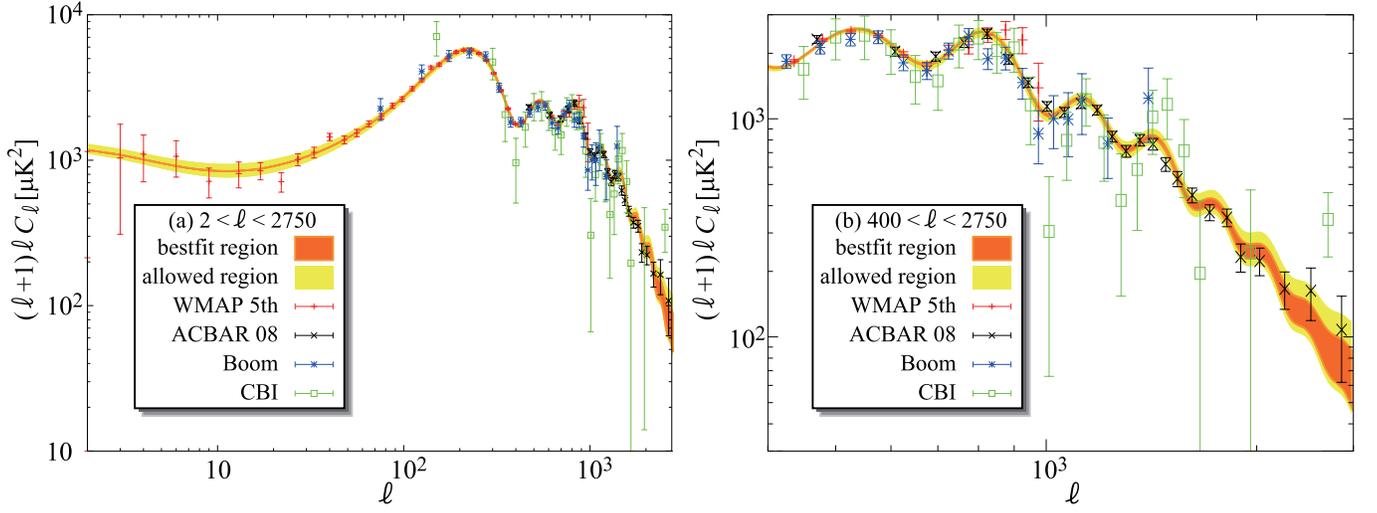}
\caption{\label{fig10}
Comparison of the best-fit computed total power spectrum with the observed CMB spectra.  
Plots show various ranges for: (a) TT($2 < \ell < 2750$) and (b) TT($400 < \ell < 2750$) modes.
Orange regions are from the best-fit parameter set
and allowed regions are from constrained parameter set as Table.1. 
Red, black, blue, green, orange, and purple dots with error bars show 
WMAP 5yr, ACBAR 08, Boomerang and CBI data, respectively. 
Since the SZ effect depends upon frequencies of the CMB for the TT mode, we plot the best fit and allowed regions in panels (a) and (b) surrounded by the curves with the SZ effect at the K(22.8GHz) band (upper curves) and without the SZ effect(lower curves).
}
\end{figure*}

\begin{figure*}[h]
\includegraphics[width=1.0\textwidth]{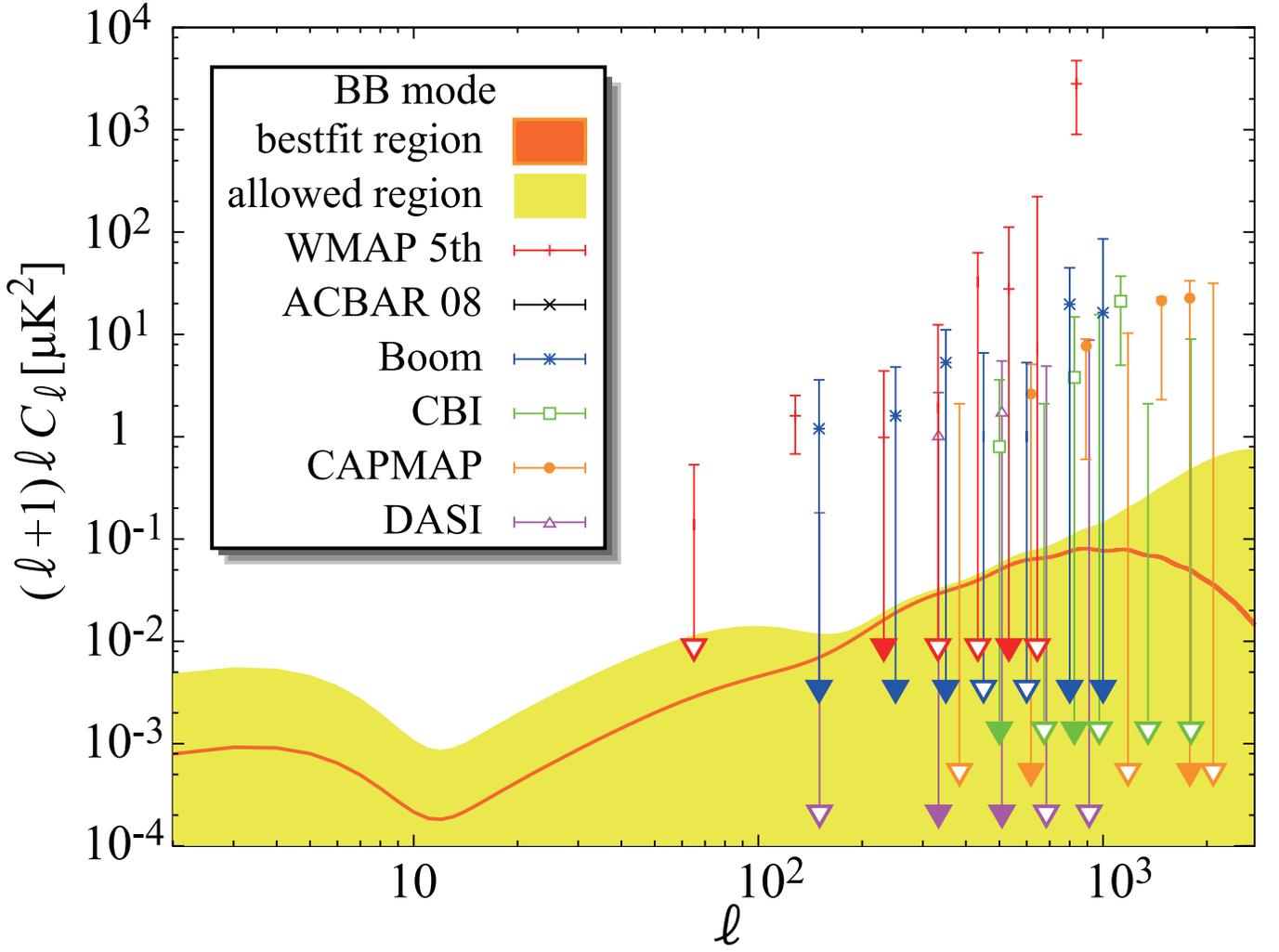}
\caption{\label{fig11} 
Comparison of the best-fit computed total power spectrum with the observed CMB spectra.  
The  orange region is from the best-fit parameter set
and the  allowed region is from the constrained parameter set as given in Table.1. 
Red, black, blue, green, orange, and purple dots with error bars show 
WMAP 5yr, ACBAR 08, Boomerang, CBI, CAPMAP, and DASI data, respectively. 
Downward arrows for the error bars on this figure indicate that the data points are upper limits.
Since the BB mode is not affected by the SZ effect, the upper curves and lower curves in this figure are defined by the constrained cosmological parameters of Table.1.
}
\end{figure*}

A PMF affects not only the temperature fluctuations, but also the polarization of the CMB.  
Although we fit all available polarization data, it turns out that the TT and BB modes (where T is the temperature fluctuation and B is the curl-like component of polarization) are the most important.
  Figures \ref{fig10} and \ref{fig11} show a comparison of the computed best-fit total power spectrum with the observed CMB spectrum.  Plots show various spectra for the TT and  BB modes.
We plot  the best fit and allowed regions both including  the SZ effect (scattering from re-ionized electrons) at the K(22.8GHz) band (upper curves) and without the SZ effect (lower curves) in Figure\ref{fig9}.  Including the SZ effect only slightly diminishes the best fit magnitude of the PMF.
Although the CBI point falls about 1 $\sigma$ above the best fit, the $\chi^2$  is  dominated  by the better ACBAR08 data and this point does not significantly affect the deduced PMF parameters.
\begin{table}
\begin{tabular}{ccc} 
\multicolumn{3}{c}{\it Cosmological Parameters}\\
\hline
\multicolumn{1}{c}{Parameter} &
\multicolumn{1}{c}{
mean}&
\multicolumn{1}{c}{best fit}\\
\hline
$\Omega_b h^2$ &
$0.02320 \pm 0.00059$ &
$0.02295$ \\
$\Omega_c h^2$ &
$0.1094\pm0.0046$ &
$0.1093$ \\
$\tau_C$ &
$0.087\pm0.017$ &
$0.082$ \\
$n_s$ &
$0.977\pm0.016$ &
$0.970$ \\
$\ln(10^{10}A_s)$ &
$3.07\pm0.036$ &
$ 3.06$\\
$A_t/A_s$ &
$< 0.170 ( 68\% \mathrm{CL}), < 0.271 ( 95\% \mathrm{CL})$ &
$0.0088$\\
$|B_\lambda|\mathrm{(nG)}$ &
$\mathbf{< 2.10} ( 68\% \mathrm{CL}), \mathbf{< 2.98} ( 95\% \mathrm{CL})$ &
$\mathbf{0.85}$\\
$n_\mathrm{B}$ &
$\mathbf{< -1.19} ( 68\% \mathrm{CL}), \mathbf{< -0.25} ( 95\% \mathrm{CL})$ &
$\mathbf{-2.37}$\\
\hline
$\sigma_8$ &
$0.812^{+0.028}_{-0.033}$ &
$ 0.794$\\
$H_0$ &
$73.3\pm2.2$ &
$ 72.8$\\
$z_\mathrm{reion}$ &
$10.9\pm 1.4$ &
$ 10.5$\\
$\mathrm{Age(Gyr)}$ &
$13.57\pm 0.12$ &
$ 13.62$\\
\hline\hline
\end{tabular}
\caption{PMF parameters and $\Lambda$CDM model parameters and 68\%
confidence intervals ($A_t/A_s$  is a 95\% CL) from a fit to the WMAP\cite{Hinshaw:2008kr} + ACBAR\cite{Kuo:2006ya} + CBI\cite{Sievers:2005gj} +Boomerang\cite{Jones:2005yb} +  2dFDR\cite{Cole:2005sx} data. }
\end{table}
\section{Discussion}
Effects of a PMF on the early universe have  been studied by a number  of researchers including ourselves.  A summary of some of the relevant topics and their implications for different  PMF field strengths is given in Table 2.  For example, in the first row, if a PMF  originated during a post inflation epoch, its strength is probably $|\mathrm{B}_\lambda| \ll 1 nG$, while if it was generated during inflation its field strength  at the surface of last photon scattering was likely $|\mathrm{B}_\lambda| \sim 1 nG$. 
Furthermore the BB mode of the CMB from the tensor component is strongly affected by the gravitational wave background. Therefore, the BB mode also shows characteristics of the gravitational wave background from the PMF and an inflation origin.  If the PMF has a field strength order of $1~nG$, then from figure \ref{fig3} and \ref{fig4}, we expect that the gravity wave background  (without the PMF) dominates the BB mode for smaller $\ell < 200$, while, the PMF dominates this mode for $\ell > 200$. Hence, if the gravity wave background without the PMF is generated by inflation, the origin of the gravity wave background on large scales will be inflation, while  the background for smaller scales is from the PMF.

Based upon the current status of the field, we can summarize the following:
 1) A PMF mainly affects the CMB temperature fluctuations and polarization anisotropies for small angular scales; 2)  Fits to observations of the CMB polarization anisotropies for small angular scales are generally better when effects of the PMF are included; and  
3)  A PMF can either increase or decrease the matter density fluctuations on scales  less than that  of galaxy clusters.
\begin{table}[b]
\begin{tabular}{|m{10em}||m{17em}|m{17em}|} 
\hline
\multicolumn{1}{|c||}{Topics} &
\multicolumn{1}{c}{$|\mathrm{B}_\lambda| \ll 1 nG$}&
\multicolumn{1}{|c|}{$|\mathrm{B}_\lambda| \sim 1 nG$}\\
\hline
\hline
the origin of PMF &
post-inflation &
Inflation \\
\hline
the gravitational wave background &
Inflation &
Large scaleFinflation 
Small scaleFPMF 
\\
\hline
the formation of LSS &
necessity of some generation model for the magnetic field or  vorticity  &
necessity for proper initial conditions considering the magnetic field and vorticity \\
\hline
the magnetic field in clusters of galaxies &
necessity of some global magnetic field generation mode + amplification model  &
PMF origin \\
\hline
the re-ionization and the first objects &
--- &
Formation of the  first stars and the re-ionization are stimulated by a PMF.\\
\hline
\end{tabular}
\caption{Effects of the PMF on each physical processes}
\end{table}
\subsection{Comparison with other determinations}
As noted above, in our previous numerical estimate \cite{Yamazaki:2007oc},  the curvature perturbations of the scalar mode  were too small to stabilize the numerical calculations for large scales and early times. 
This is caused by an unwanted cancellation between contributions from the  primordial magnetic field and the primordial radiation fields which seemed to imply excess power for the lowest mulitpoles of the CMB.
In order to solve this problem, we have adopted the  numerical approach  and  initial conditions for the matter contributions from    Ref.~\cite{Kojima:2009ms}.  This formulation  has been developed independently and gives consistent results for large scales and early  times. 
We have now confirmed that our results are consistent not only with results of Ref.~\cite{Kojima:2009ms}, but also those of  other groups which employ semi-numerical methods [e.g.~Ref.~\cite{Paoletti:2008ck}]. 
\subsection{Influence of the PMF on Large Scale Structure}
In Ref.~\cite{Yamazaki:2009na} we deduced  the effects of a PMF on the  matter energy density fields by considering a stochastic PMF that depends upon scale.   We then  quantitatively discussed the effect of a PMF on the seeds of LSS in the early universe.
We have also considered more general effects of the PMF than those considered in  previous work. 
For example, we included not only the magnetic field tension but also the increases in pressure and energy
density perturbations from the field.
Furthermore, by considering the correlation between the PMF and the matter density fluctuations, and utilizing  mathematically exact stochastic PMF power spectrum sets,  we have obtained reasonable and accurate models for the evolution of baryons, CDM, and photons, and therefore, the large scale structure. 
We have shown \cite{Yamazaki:2009na} that the PMF can play a very  different role on the evolution of density perturbations depending upon  how the PMF and matter fluctuations are  correlated. 
We have also considered the fact that after  decoupling, the CDM is  influenced indirectly by the PMF through gravitational interaction.  

We reported in Ref.~ \cite{Yamazaki:2007oc}  that a PMF on small angular scales scales
($\lambda \approx$1Mpc) and a field strength of  $B_\lambda \sim 1$ nG 
\cite{Mack:2001gc,Lewis:2004ef,Yamazaki:2004vq,Yamazaki:2006bq,Yamazaki:2006ah,Durrer:1999bk} provides a new interpretation for the excess of CMB anisotropies for high multipoles. We showed that if a PMF with such strength was present, it is very likely that it has affected the formation of large scale structure.
In Ref.~\cite{Yoshida:2003sy} it was suggested that in order to avoid false coupling between  the baryons and the CDM on small scales, one should use  independent transfer functions for the baryons and CDM. 
A PMF would be another source of this difference in the transfer function for baryons and the CDM. 
Since the density perturbations in the early universe have evolved to the present LSS, the evolution of the LSS with a PMF becomes  different from that without a PMF. 
We have shown that the baryon and CDM energy density perturbations evolve  very differently
 in the presence of a PMF;  and with the PMF taken into consideration, the evolution of large scale structure  becomes more complicated.
\subsection{Future Observations}
Magnetic fields affect the  generation and evolution of a wide variety  of cosmological and stellar objects in the universe, e.g. galaxy clusters, galaxies,  primordial stars, etc.,  (cf.~Table 2).
Therefore, it is important to constrain the parameters characterizing the  PMF  in order to study their effects on physical processes in the universe.
In this regard, forthcoming   missions, e.g. QUIET, PolarBear, and Planck are expected  to provide much more precise data on the CMB temperature and polarization anisotropies.   At the same time data on the formation of large scale structure will continue to be  measured more precisely both by future space-based observations and from  continuing ground-based observations such as the SDSS and 2dF projects.

As summarized in this review, at the present time, we now understand the   main effects of a  PMF  on both the CMB and the matter power spectrum.  The important task remaining in the future   will be to apply our new and correct numerical methods for  calculating all PMF fluctuations of the scalar, vector and tensor modes in the TT and BB power spectra (see figures \ref{fig9}, \ref{fig10} and \ref{fig11}) to fit the new observations as they emerge.   Only through a critical comparison with such future data can  we hope to ultimately clarify the  magnitude and origin  of the PMF in the early universe. 

Although, some degeneracy between the standard cosmological  and PMF parameters is expected in the present data, we believe that by simultaneously  investigating both the  CMB anisotropies and matter power spectra one can  resolve these degeneracies.
The ultimate establishment of the existence and properties of a PMF will permit a better understanding of its  generation and evolution and also provide new insight into the formation of LSS as well as a possible new probe of the physics of the early universe.
\begin{acknowledgments}
This work has been supported in part by Grants-in-Aid for Scientific
Research (20105004, 20244035, 21740177, 20169444) of the Ministry of Education, Culture, Sports,
Science and Technology of Japan.  This
work is also supported by the JSPS Core-to-Core Program, International
Research Network for Exotic Femto Systems (EFES).  Work at UND supported in part by the US Department of Energy under research grant DE-FG02-95-ER40934.
\end{acknowledgments}
\appendix
\section{Initial conditions}
In this Appendix, we introduce the initial conditions of compensated magnetic modes with massless neutrinos in the synchronous gauge which were derived by Ref.\cite{ Shaw:2009nf}.
Subscripts in this appendix indicate the standard cosmological species: 
$\gamma$: photon,
$\nu$: massless neutrino,
$\mathrm{b}$: baryon, and
$\mathrm{CDM}$: cold dark matter.
We define the density closure parameters as the ratios of density parameter as $\Omega_x = \rho_x/\rho_\mathrm{cr}$,
where $\rho_\mathrm{cr}$ is the critical density of the universe and the subscript $x$ denotes the various species listed above. 
We also define ratios of 
$R_\gamma = \Omega_\gamma/\Omega_\mathrm{r}$,
$R_\nu = \Omega_\nu/\Omega_\mathrm{r}$, and 
$R_\mathrm{b} = \Omega_\mathrm{b}/\Omega_\mathrm{m}$,
where 
$\Omega_\mathrm{r} = \Omega_\gamma + \Omega_\nu$ and 
$\Omega_\mathrm{m} = \Omega_\mathrm{CDM} + \Omega_\mathrm{b}$.
\center{$scalar~mode$}
\begin{eqnarray}
&&\mathrm{Potential}\nonumber\\
h
	&=&
	-3
	R_\gamma
	\frac{\beta}{\alpha}
	\tau
	+
	\frac{9}{2}
	R_\gamma
	\frac{\beta^2}{\alpha^2}
	\tau^2
	+\mathcal{O}(\tau^3)
	\\
&&\mathrm{Curvature~potential}\nonumber\\
\eta
	&=&
	\frac{1}{2}
	R_\gamma
	\frac{\beta}{\alpha}
	\tau
	+
	\left[
		\frac{R_\nu R_\gamma}
		     {6(4R_\nu+15)}
		     k^2
		-
		\frac{3}{4}
		R_\gamma
		\frac{\beta^2}{\alpha^2}
	\right]
	\tau^2
	+\mathcal{O}(\tau^3)
	\\
&&\mathrm{Density~perturbations}\nonumber\\
\delta^\mathrm{(S)}_\mathrm{CDM}
	&=&
	-
	\frac{3}{4}
	R_\gamma
	+
	\frac{3}{2}
	R_\gamma
	\frac{\beta}{\alpha}
	\tau
	-
	\frac{9}{4}
	R_\gamma
	\frac{\beta^2}{\alpha^2}
	\tau^2
	+\mathcal{O}(\tau^3)
	\\
\delta^\mathrm{(S)}_\gamma
	&=&
	-
	R_\gamma
	+
	2
	R_\gamma
	\frac{\beta}{\alpha}
	\tau
	-
	\left(
		\frac{R_\nu}{6}
		k^3
		+
		3
		R_\gamma
		\frac{\beta^2}{\alpha^2}
	\right)
	\tau^2
	+\mathcal{O}(\tau^3)
	\\
\delta^\mathrm{(S)}_\nu
	&=&
	-
	R_\gamma
	+
	2
	R_\gamma
	\frac{\beta}{\alpha}
	\tau
	+
	\left(
		\frac{R_\nu}{6}
		k^3
		-
		3
		R_\gamma
		\frac{\beta^2}{\alpha^2}
	\right)
	\tau^2
	+\mathcal{O}(\tau^3)
	\\
\delta^\mathrm{(S)}_\mathrm{b}
	&=&
	\frac{3}{4}
	\delta^\mathrm{(S)}_\gamma
\\
&&\mathrm{Velocity~perturbations}\nonumber\\
v^\mathrm{(S)}_\gamma
	&=&
	\frac{1}{4}
	R_\nu
	k\tau
	+
	 \frac{1}{4}
	 (1-2R_\nu-3R_\mathrm{b}R_\nu+R^2_\nu)
	 \frac{\beta}{\alpha}
	 k\tau^2
	+\mathcal{O}(\tau^3)
	\\
v^\mathrm{(S)}_\nu
	&=&
	\frac{1}{4}
	R_\gamma
	k\tau
	+
	 \frac{1}{4}
	 R_\gamma
	 \frac{\beta}{\alpha}
	 k\tau^2
	+\mathcal{O}(\tau^3)
	\\
v^\mathrm{(S)}_\mathrm{b}
	&=&
    v^\mathrm{(S)}_\gamma
\\
&&\mathrm{Anisotropic stress}\nonumber\\
\Pi^\mathrm{(S)}_\nu
	&=&
	-\frac{3R_\gamma}{4R_\nu + 15}
	 k^2\tau^2
	+\mathcal{O}(\tau^3)
    \\
&&\mathrm{Boltzmann~hierarchy~of~neutrinos}\nonumber\\
F^\mathrm{(S)}_{\nu 3}
	&=&
	-\frac{R_\gamma}{4R_\nu + 15}
	 k^3\tau^3
	+\mathcal{O}(\tau^4)
\end{eqnarray}
\center{$vector~mode$}
\begin{eqnarray}
&&\mathrm{Vector~potential}\nonumber\\
\sigma^\mathrm{(V)}
	&=&
	\frac{15R_\gamma}{56R_\nu + 210}
	k\tau
	-
	\frac{225R_\gamma}
	     {7(8R^2_\nu + 90R_\nu + 225)}
	\frac{\beta}{\alpha}
	k\tau^2
	\nonumber\\
	&& +\mathcal{O}(\tau^3)
	\\
&&\mathrm{Velocity~perturbations}\nonumber\\
v^\mathrm{(V)}_\gamma
	&=&
	-\frac{1}{8}k\tau
	+
	 \frac{3R_\mathrm{b}}{8R_\gamma}
	 \frac{\beta}{\alpha}
	 k\tau^2
	+\mathcal{O}(\tau^3)
	\\
v^\mathrm{(V)}_\nu
	&=&
	-\frac{R_\gamma}{8R_\nu}k\tau
	+\mathcal{O}(\tau^3)
	\\
v^\mathrm{(V)}_\mathrm{b}
	&=&
	v^\mathrm{(V)}_\gamma
	\\
&&\mathrm{Anisotropic stress}\nonumber\\
\Pi^\mathrm{(V)}_\nu
	&=&
	 -\frac{R_\gamma}{R_\nu}
	+
	\frac{45R_\gamma}
	     {56R^2_\nu + 210R_\nu}
	 k^2\tau^2
	+\mathcal{O}(\tau^3)
    \\
&&\mathrm{Boltzmann~hierarchy~of~neutrinos}\nonumber\\
F^\mathrm{(V)}_{\nu 3}
	&=&
	-\sqrt{\frac{2}{3}}
	 \frac{R_\gamma}{R_\nu}
	 k\tau
	+\mathcal{O}(\tau^3)
\end{eqnarray}
\center{$tensor~mode$}
\begin{eqnarray}
&&\mathrm{tensor~potential}\nonumber\\
\mathcal{H}^\mathrm{(T)}
	&=&
	\frac{5R_\gamma}{28(4R_\nu + 15)}
	k^2\tau^2
	+\mathcal{O}(\tau^3)
	\\
&&\mathrm{Anisotropic stress}\nonumber\\
\Pi^\mathrm{(T)}_\nu
	&=&
	-\frac{R_\gamma}{R_\nu}
	+
	\frac{15R_\gamma}
	     {56R^2_\nu + 210R_\nu}
	 k^2\tau^2
	+\mathcal{O}(\tau^3)
	\\
&&\mathrm{Boltzmann~hierarchy~of~neutrinos}\nonumber\\
F^\mathrm{(T)}_{\nu 3}
	&=&
	-\frac{\sqrt{5}}{2}
	 \frac{R_\gamma}{R_\nu}
	 k\tau
	+\mathcal{O}(\tau^3)
\end{eqnarray}
\newpage

\begin{thebibliography}{81}
\expandafter\ifx\csname natexlab\endcsname\relax\def\natexlab#1{#1}\fi
\expandafter\ifx\csname bibnamefont\endcsname\relax
  \def\bibnamefont#1{#1}\fi
\expandafter\ifx\csname bibfnamefont\endcsname\relax
  \def\bibfnamefont#1{#1}\fi
\expandafter\ifx\csname citenamefont\endcsname\relax
  \def\citenamefont#1{#1}\fi
\expandafter\ifx\csname url\endcsname\relax
  \def\url#1{\texttt{#1}}\fi
\expandafter\ifx\csname urlprefix\endcsname\relax\def\urlprefix{URL }\fi
\providecommand{\bibinfo}[2]{#2}
\providecommand{\eprint}[2][]{\url{#2}}

\bibitem[{\citenamefont{Wolfe et~al.}(1992)\citenamefont{Wolfe, Lanzetta, and
  Oren}}]{Wolfe:1992ab}
\bibinfo{author}{\bibfnamefont{A.~M.} \bibnamefont{Wolfe}},
  \bibinfo{author}{\bibfnamefont{K.~M.} \bibnamefont{Lanzetta}},
  \bibnamefont{and} \bibinfo{author}{\bibfnamefont{A.~L.} \bibnamefont{Oren}},
  \bibinfo{journal}{ApJ.} \textbf{\bibinfo{volume}{388}}, \bibinfo{pages}{17}
  (\bibinfo{year}{1992}).

\bibitem[{\citenamefont{Clarke et~al.}(2001)\citenamefont{Clarke, Kronberg, and
  Boehringer}}]{Clarke:2000bz}
\bibinfo{author}{\bibfnamefont{T.~E.} \bibnamefont{Clarke}},
  \bibinfo{author}{\bibfnamefont{P.~P.} \bibnamefont{Kronberg}},
  \bibnamefont{and}
  \bibinfo{author}{\bibfnamefont{H.}~\bibnamefont{Boehringer}},
  \bibinfo{journal}{Astrophys. J.} \textbf{\bibinfo{volume}{547}},
  \bibinfo{pages}{L111} (\bibinfo{year}{2001}).

\bibitem[{\citenamefont{Xu et~al.}(2006)\citenamefont{Xu, Kronberg, Habib, and
  Dufton}}]{Xu:2005rb}
\bibinfo{author}{\bibfnamefont{Y.}~\bibnamefont{Xu}},
  \bibinfo{author}{\bibfnamefont{P.~P.} \bibnamefont{Kronberg}},
  \bibinfo{author}{\bibfnamefont{S.}~\bibnamefont{Habib}}, \bibnamefont{and}
  \bibinfo{author}{\bibfnamefont{Q.~W.} \bibnamefont{Dufton}},
  \bibinfo{journal}{Astrophys. J.} \textbf{\bibinfo{volume}{637}},
  \bibinfo{pages}{19} (\bibinfo{year}{2006}).

\bibitem[{\citenamefont{Turner and Widrow}(1988)}]{Turner:1987bw}
\bibinfo{author}{\bibfnamefont{M.~S.} \bibnamefont{Turner}} \bibnamefont{and}
  \bibinfo{author}{\bibfnamefont{L.~M.} \bibnamefont{Widrow}},
  \bibinfo{journal}{Phys. Rev.} \textbf{\bibinfo{volume}{D 37}},
  \bibinfo{pages}{2743} (\bibinfo{year}{1988}).

\bibitem[{\citenamefont{Ratra}(1992)}]{Ratra:1991bn}
\bibinfo{author}{\bibfnamefont{B.}~\bibnamefont{Ratra}},
  \bibinfo{journal}{Astrophys. J.} \textbf{\bibinfo{volume}{391}},
  \bibinfo{pages}{L1} (\bibinfo{year}{1992}).

\bibitem[{\citenamefont{Bamba and Yokoyama}(2004)}]{Bamba:2004cu}
\bibinfo{author}{\bibfnamefont{K.}~\bibnamefont{Bamba}} \bibnamefont{and}
  \bibinfo{author}{\bibfnamefont{J.}~\bibnamefont{Yokoyama}},
  \bibinfo{journal}{Phys. Rev.} \textbf{\bibinfo{volume}{D 70}},
  \bibinfo{pages}{083508} (\bibinfo{year}{2004}).

\bibitem[{\citenamefont{Vachaspati}(1991)}]{Vachaspati:1991nm}
\bibinfo{author}{\bibfnamefont{T.}~\bibnamefont{Vachaspati}},
  \bibinfo{journal}{Phys. Lett.} \textbf{\bibinfo{volume}{B265}},
  \bibinfo{pages}{258} (\bibinfo{year}{1991}).

\bibitem[{\citenamefont{Kibble and Vilenkin}(1995)}]{Kibble:1995aa}
\bibinfo{author}{\bibfnamefont{T.~W.~B.} \bibnamefont{Kibble}}
  \bibnamefont{and} \bibinfo{author}{\bibfnamefont{A.}~\bibnamefont{Vilenkin}},
  \bibinfo{journal}{Phys. Rev.} \textbf{\bibinfo{volume}{D52}},
  \bibinfo{pages}{679} (\bibinfo{year}{1995}).

\bibitem[{\citenamefont{Ahonen and Enqvist}(1998)}]{Ahonen:1997wh}
\bibinfo{author}{\bibfnamefont{J.}~\bibnamefont{Ahonen}} \bibnamefont{and}
  \bibinfo{author}{\bibfnamefont{K.}~\bibnamefont{Enqvist}},
  \bibinfo{journal}{Phys. Rev.} \textbf{\bibinfo{volume}{D57}},
  \bibinfo{pages}{664} (\bibinfo{year}{1998}).

\bibitem[{\citenamefont{Joyce and Shaposhnikov}(1997)}]{Joyce:1997uy}
\bibinfo{author}{\bibfnamefont{M.}~\bibnamefont{Joyce}} \bibnamefont{and}
  \bibinfo{author}{\bibfnamefont{M.~E.} \bibnamefont{Shaposhnikov}},
  \bibinfo{journal}{Phys. Rev. Lett.} \textbf{\bibinfo{volume}{79}},
  \bibinfo{pages}{1193} (\bibinfo{year}{1997}).

\bibitem[{\citenamefont{Takahashi et~al.}(2005)\citenamefont{Takahashi, Ichiki,
  Ohno, and Hanayama}}]{Takahashi:2005nd}
\bibinfo{author}{\bibfnamefont{K.}~\bibnamefont{Takahashi}},
  \bibinfo{author}{\bibfnamefont{K.}~\bibnamefont{Ichiki}},
  \bibinfo{author}{\bibfnamefont{H.}~\bibnamefont{Ohno}}, \bibnamefont{and}
  \bibinfo{author}{\bibfnamefont{H.}~\bibnamefont{Hanayama}},
  \bibinfo{journal}{Phys. Rev. Lett.} \textbf{\bibinfo{volume}{95}},
  \bibinfo{pages}{121301} (\bibinfo{year}{2005}).

\bibitem[{\citenamefont{Hanayama et~al.}(2005)}]{Hanayama:2005hd}
\bibinfo{author}{\bibfnamefont{H.}~\bibnamefont{Hanayama}}
  \bibnamefont{et~al.}, \bibinfo{journal}{Astrophys. J.}
  \textbf{\bibinfo{volume}{633}}, \bibinfo{pages}{941} (\bibinfo{year}{2005}).

\bibitem[{\citenamefont{Ichiki et~al.}(2006)\citenamefont{Ichiki, Takahashi,
  Ohno, Hanayama, and Sugiyama}}]{ichiki:2006sc}
\bibinfo{author}{\bibfnamefont{K.}~\bibnamefont{Ichiki}},
  \bibinfo{author}{\bibfnamefont{K.}~\bibnamefont{Takahashi}},
  \bibinfo{author}{\bibfnamefont{H.}~\bibnamefont{Ohno}},
  \bibinfo{author}{\bibfnamefont{H.}~\bibnamefont{Hanayama}}, \bibnamefont{and}
  \bibinfo{author}{\bibfnamefont{N.}~\bibnamefont{Sugiyama}},
  \bibinfo{journal}{Science} \textbf{\bibinfo{volume}{311}},
  \bibinfo{pages}{827} (\bibinfo{year}{2006}).

\bibitem[{\citenamefont{Subramanian and
  Barrow}(1998{\natexlab{a}})}]{Subramanian:1998fn}
\bibinfo{author}{\bibfnamefont{K.}~\bibnamefont{Subramanian}} \bibnamefont{and}
  \bibinfo{author}{\bibfnamefont{J.~D.} \bibnamefont{Barrow}},
  \bibinfo{journal}{Phys. Rev. Lett.} \textbf{\bibinfo{volume}{81}},
  \bibinfo{pages}{3575} (\bibinfo{year}{1998}{\natexlab{a}}).

\bibitem[{\citenamefont{Mack et~al.}(2002)\citenamefont{Mack, Kahniashvili, and
  Kosowsky}}]{Mack:2001gc}
\bibinfo{author}{\bibfnamefont{A.}~\bibnamefont{Mack}},
  \bibinfo{author}{\bibfnamefont{T.}~\bibnamefont{Kahniashvili}},
  \bibnamefont{and} \bibinfo{author}{\bibfnamefont{A.}~\bibnamefont{Kosowsky}},
  \bibinfo{journal}{Phys. Rev.} \textbf{\bibinfo{volume}{D 65}},
  \bibinfo{pages}{123004} (\bibinfo{year}{2002}).

\bibitem[{\citenamefont{Subramanian and Barrow}(2002)}]{Subramanian:2002nh}
\bibinfo{author}{\bibfnamefont{K.}~\bibnamefont{Subramanian}} \bibnamefont{and}
  \bibinfo{author}{\bibfnamefont{J.~D.} \bibnamefont{Barrow}},
  \bibinfo{journal}{Mon. Not. Roy. Astron. Soc.}
  \textbf{\bibinfo{volume}{335}}, \bibinfo{pages}{L57} (\bibinfo{year}{2002}).

\bibitem[{\citenamefont{Lewis}(2004)}]{Lewis:2004ef}
\bibinfo{author}{\bibfnamefont{A.}~\bibnamefont{Lewis}},
  \bibinfo{journal}{Phys. Rev.} \textbf{\bibinfo{volume}{D 70}},
  \bibinfo{pages}{043011} (\bibinfo{year}{2004}).

\bibitem[{\citenamefont{Yamazaki
  et~al.}(2005{\natexlab{a}})\citenamefont{Yamazaki, Ichiki, and
  Kajino}}]{Yamazaki:2004vq}
\bibinfo{author}{\bibfnamefont{D.~G.} \bibnamefont{Yamazaki}},
  \bibinfo{author}{\bibfnamefont{K.}~\bibnamefont{Ichiki}}, \bibnamefont{and}
  \bibinfo{author}{\bibfnamefont{T.}~\bibnamefont{Kajino}},
  \bibinfo{journal}{Astrophys. J.} \textbf{\bibinfo{volume}{625}},
  \bibinfo{pages}{L1} (\bibinfo{year}{2005}{\natexlab{a}}).

\bibitem[{\citenamefont{Kahniashvili and Ratra}(2005)}]{Kahniashvili:2005xe}
\bibinfo{author}{\bibfnamefont{T.}~\bibnamefont{Kahniashvili}}
  \bibnamefont{and} \bibinfo{author}{\bibfnamefont{B.}~\bibnamefont{Ratra}},
  \bibinfo{journal}{Phys. Rev.} \textbf{\bibinfo{volume}{D 71}},
  \bibinfo{pages}{103006} (\bibinfo{year}{2005}).

\bibitem[{\citenamefont{Challinor}(2004)}]{Challinor:2005ye}
\bibinfo{author}{\bibfnamefont{A.}~\bibnamefont{Challinor}},
  \bibinfo{journal}{Lect. Notes Phys.} \textbf{\bibinfo{volume}{653}},
  \bibinfo{pages}{71} (\bibinfo{year}{2004}).

\bibitem[{\citenamefont{Dolgov}(2005)}]{Dolgov:2005ti}
\bibinfo{author}{\bibfnamefont{A.~D.} \bibnamefont{Dolgov}}
  (\bibinfo{year}{2005}), \eprint{astro-ph/0503447}.

\bibitem[{\citenamefont{Gopal and Sethi}(2005)}]{Gopal:2005sg}
\bibinfo{author}{\bibfnamefont{R.}~\bibnamefont{Gopal}} \bibnamefont{and}
  \bibinfo{author}{\bibfnamefont{S.~K.} \bibnamefont{Sethi}},
  \bibinfo{journal}{Phys. Rev.} \textbf{\bibinfo{volume}{D 72}},
  \bibinfo{pages}{103003} (\bibinfo{year}{2005}).

\bibitem[{\citenamefont{Yamazaki
  et~al.}(2005{\natexlab{b}})\citenamefont{Yamazaki, Ichiki, and
  Kajino}}]{Yamazaki:2005yd}
\bibinfo{author}{\bibfnamefont{D.~G.} \bibnamefont{Yamazaki}},
  \bibinfo{author}{\bibfnamefont{K.}~\bibnamefont{Ichiki}}, \bibnamefont{and}
  \bibinfo{author}{\bibfnamefont{T.}~\bibnamefont{Kajino}},
  \bibinfo{journal}{Nuclear Physics A} \textbf{\bibinfo{volume}{758}},
  \bibinfo{pages}{791} (\bibinfo{year}{2005}{\natexlab{b}}).

\bibitem[{\citenamefont{Kahniashvili and Ratra}(2007)}]{Kahniashvili:2006hy}
\bibinfo{author}{\bibfnamefont{T.}~\bibnamefont{Kahniashvili}}
  \bibnamefont{and} \bibinfo{author}{\bibfnamefont{B.}~\bibnamefont{Ratra}},
  \bibinfo{journal}{Phys. Rev.} \textbf{\bibinfo{volume}{D75}},
  \bibinfo{pages}{023002} (\bibinfo{year}{2007}).

\bibitem[{\citenamefont{Yamazaki
  et~al.}(2006{\natexlab{a}})\citenamefont{Yamazaki, Ichiki, Umezu, and
  Hanayama}}]{Yamazaki:2006mi}
\bibinfo{author}{\bibfnamefont{D.~G.} \bibnamefont{Yamazaki}},
  \bibinfo{author}{\bibfnamefont{K.}~\bibnamefont{Ichiki}},
  \bibinfo{author}{\bibfnamefont{K.~I.} \bibnamefont{Umezu}}, \bibnamefont{and}
  \bibinfo{author}{\bibfnamefont{H.}~\bibnamefont{Hanayama}},
  \bibinfo{journal}{Phys. Rev.} \textbf{\bibinfo{volume}{D 74}},
  \bibinfo{pages}{123518} (\bibinfo{year}{2006}{\natexlab{a}}).

\bibitem[{\citenamefont{Yamazaki
  et~al.}(2006{\natexlab{b}})\citenamefont{Yamazaki, Ichiki, Kajino, and
  Mathews}}]{Yamazaki:2006bq}
\bibinfo{author}{\bibfnamefont{D.~G.} \bibnamefont{Yamazaki}},
  \bibinfo{author}{\bibfnamefont{K.}~\bibnamefont{Ichiki}},
  \bibinfo{author}{\bibfnamefont{T.}~\bibnamefont{Kajino}}, \bibnamefont{and}
  \bibinfo{author}{\bibfnamefont{G.~J.} \bibnamefont{Mathews}},
  \bibinfo{journal}{Astrophys. J.} \textbf{\bibinfo{volume}{646}},
  \bibinfo{pages}{719} (\bibinfo{year}{2006}{\natexlab{b}}).

\bibitem[{\citenamefont{Yamazaki
  et~al.}(2006{\natexlab{c}})\citenamefont{Yamazaki, Ichiki, Kajino, and
  Mathews}}]{Yamazaki:2006ah}
\bibinfo{author}{\bibfnamefont{D.~G.} \bibnamefont{Yamazaki}},
  \bibinfo{author}{\bibfnamefont{K.}~\bibnamefont{Ichiki}},
  \bibinfo{author}{\bibfnamefont{T.}~\bibnamefont{Kajino}}, \bibnamefont{and}
  \bibinfo{author}{\bibfnamefont{G.~J.} \bibnamefont{Mathews}},
  \bibinfo{journal}{PoS(NIC-IX).} p. \bibinfo{pages}{194}
  (\bibinfo{year}{2006}{\natexlab{c}}).

\bibitem[{\citenamefont{Giovannini}(2006)}]{Giovannini:2006kc}
\bibinfo{author}{\bibfnamefont{M.}~\bibnamefont{Giovannini}},
  \bibinfo{journal}{Phys. Rev.} \textbf{\bibinfo{volume}{D 74}},
  \bibinfo{pages}{063002} (\bibinfo{year}{2006}).

\bibitem[{\citenamefont{{Yamazaki} et~al.}(2008)\citenamefont{{Yamazaki},
  {Ichiki}, {Kajino}, and {Mathews}}}]{Yamazaki:2007oc}
\bibinfo{author}{\bibfnamefont{D.~G.} \bibnamefont{{Yamazaki}}},
  \bibinfo{author}{\bibfnamefont{K.}~\bibnamefont{{Ichiki}}},
  \bibinfo{author}{\bibfnamefont{T.}~\bibnamefont{{Kajino}}}, \bibnamefont{and}
  \bibinfo{author}{\bibfnamefont{G.~J.} \bibnamefont{{Mathews}}},
  \bibinfo{journal}{Phys. Rev.} \textbf{\bibinfo{volume}{D 77}},
  \bibinfo{pages}{043005} (\bibinfo{year}{2008}).

\bibitem[{\citenamefont{Paoletti et~al.}(2009)\citenamefont{Paoletti, Finelli,
  and Paci}}]{Paoletti:2008ck}
\bibinfo{author}{\bibfnamefont{D.}~\bibnamefont{Paoletti}},
  \bibinfo{author}{\bibfnamefont{F.}~\bibnamefont{Finelli}}, \bibnamefont{and}
  \bibinfo{author}{\bibfnamefont{F.}~\bibnamefont{Paci}},
  \bibinfo{journal}{Mon. Not. Roy. Astron. Soc.}
  \textbf{\bibinfo{volume}{396}}, \bibinfo{pages}{523} (\bibinfo{year}{2009}).

\bibitem[{\citenamefont{Finelli et~al.}(2008)\citenamefont{Finelli, Paci, and
  Paoletti}}]{Finelli:2008xh}
\bibinfo{author}{\bibfnamefont{F.}~\bibnamefont{Finelli}},
  \bibinfo{author}{\bibfnamefont{F.}~\bibnamefont{Paci}}, \bibnamefont{and}
  \bibinfo{author}{\bibfnamefont{D.}~\bibnamefont{Paoletti}},
  \bibinfo{journal}{Phys. Rev.} \textbf{\bibinfo{volume}{D78}},
  \bibinfo{pages}{023510} (\bibinfo{year}{2008}), \eprint{0803.1246}.

\bibitem[{\citenamefont{Yamazaki et~al.}(2008)\citenamefont{Yamazaki, Ichiki,
  Kajino, and Mathews}}]{Yamazaki:2008bb}
\bibinfo{author}{\bibfnamefont{D.~G.} \bibnamefont{Yamazaki}},
  \bibinfo{author}{\bibfnamefont{K.}~\bibnamefont{Ichiki}},
  \bibinfo{author}{\bibfnamefont{T.}~\bibnamefont{Kajino}}, \bibnamefont{and}
  \bibinfo{author}{\bibfnamefont{G.~J.} \bibnamefont{Mathews}},
  \bibinfo{journal}{Phys. Rev.} \textbf{\bibinfo{volume}{D 78}},
  \bibinfo{pages}{123001} (\bibinfo{year}{2008}).

\bibitem[{\citenamefont{{Yamazaki} et~al.}(2008)\citenamefont{{Yamazaki},
  {Ichiki}, {Kajino}, and {Mathews}}}]{2008nuco.confE.239Y}
\bibinfo{author}{\bibfnamefont{D.~G.} \bibnamefont{{Yamazaki}}},
  \bibinfo{author}{\bibfnamefont{K.}~\bibnamefont{{Ichiki}}},
  \bibinfo{author}{\bibfnamefont{T.}~\bibnamefont{{Kajino}}}, \bibnamefont{and}
  \bibinfo{author}{\bibfnamefont{G.~J.} \bibnamefont{{Mathews}}},
  \bibinfo{journal}{PoS(NIC-X).} p. \bibinfo{pages}{239}
  (\bibinfo{year}{2008}).

\bibitem[{\citenamefont{Sethi et~al.}(2008)\citenamefont{Sethi, Nath, and
  Subramanian}}]{Sethi:2008eq}
\bibinfo{author}{\bibfnamefont{S.~K.} \bibnamefont{Sethi}},
  \bibinfo{author}{\bibfnamefont{B.~B.} \bibnamefont{Nath}}, \bibnamefont{and}
  \bibinfo{author}{\bibfnamefont{K.}~\bibnamefont{Subramanian}},
  \bibinfo{journal}{Mon. Not. Roy. Astron. Soc.}
  \textbf{\bibinfo{volume}{387}}, \bibinfo{pages}{1589} (\bibinfo{year}{2008}).

\bibitem[{\citenamefont{Kojima et~al.}(2008)\citenamefont{Kojima, Ichiki,
  Yamazaki, Kajino, and Mathews}}]{Kojima:2008rf}
\bibinfo{author}{\bibfnamefont{K.}~\bibnamefont{Kojima}},
  \bibinfo{author}{\bibfnamefont{K.}~\bibnamefont{Ichiki}},
  \bibinfo{author}{\bibfnamefont{D.~G.} \bibnamefont{Yamazaki}},
  \bibinfo{author}{\bibfnamefont{T.}~\bibnamefont{Kajino}}, \bibnamefont{and}
  \bibinfo{author}{\bibfnamefont{G.~J.} \bibnamefont{Mathews}},
  \bibinfo{journal}{Phys. Rev.} \textbf{\bibinfo{volume}{D78}},
  \bibinfo{pages}{045010} (\bibinfo{year}{2008}).

\bibitem[{\citenamefont{{Kahniashvili}
  et~al.}(2008)\citenamefont{{Kahniashvili}, {Lavrelashvili}, and
  {Ratra}}}]{2008PhRvD..78f3012K}
\bibinfo{author}{\bibfnamefont{T.}~\bibnamefont{{Kahniashvili}}},
  \bibinfo{author}{\bibfnamefont{G.}~\bibnamefont{{Lavrelashvili}}},
  \bibnamefont{and} \bibinfo{author}{\bibfnamefont{B.}~\bibnamefont{{Ratra}}},
  \bibinfo{journal}{Phys. Rev.} \textbf{\bibinfo{volume}{D78}},
  \bibinfo{pages}{063012} (\bibinfo{year}{2008}).

\bibitem[{\citenamefont{Giovannini and Kunze}(2008)}]{Giovannini:2008aa}
\bibinfo{author}{\bibfnamefont{M.}~\bibnamefont{Giovannini}} \bibnamefont{and}
  \bibinfo{author}{\bibfnamefont{K.~E.} \bibnamefont{Kunze}},
  \bibinfo{journal}{Phys. Rev.} \textbf{\bibinfo{volume}{D78}},
  \bibinfo{pages}{023010} (\bibinfo{year}{2008}), \eprint{0804.3380}.

\bibitem[{\citenamefont{Yamazaki et~al.}(2010)\citenamefont{Yamazaki, Ichiki,
  Kajino, and Mathews}}]{Yamazaki:2009na}
\bibinfo{author}{\bibfnamefont{D.~G.} \bibnamefont{Yamazaki}},
  \bibinfo{author}{\bibfnamefont{K.}~\bibnamefont{Ichiki}},
  \bibinfo{author}{\bibfnamefont{T.}~\bibnamefont{Kajino}}, \bibnamefont{and}
  \bibinfo{author}{\bibfnamefont{G.~J.} \bibnamefont{Mathews}},
  \bibinfo{journal}{Phys. Rev.} \textbf{\bibinfo{volume}{D 81}},
  \bibinfo{pages}{023008} (\bibinfo{year}{2010}).

\bibitem[{\citenamefont{{Yamazaki}
  et~al.}(2010{\natexlab{a}})\citenamefont{{Yamazaki}, {Ichiki}, {Kajino}, and
  {Mathews}}}]{2010PhRvD..81j3519Y}
\bibinfo{author}{\bibfnamefont{D.~G.} \bibnamefont{{Yamazaki}}},
  \bibinfo{author}{\bibfnamefont{K.}~\bibnamefont{{Ichiki}}},
  \bibinfo{author}{\bibfnamefont{T.}~\bibnamefont{{Kajino}}}, \bibnamefont{and}
  \bibinfo{author}{\bibfnamefont{G.~J.} \bibnamefont{{Mathews}}},
  \bibinfo{journal}{Phys. Rev.} \textbf{\bibinfo{volume}{D 81}},
  \bibinfo{pages}{103519} (\bibinfo{year}{2010}{\natexlab{a}}).

\bibitem[{\citenamefont{{Yamazaki}
  et~al.}(2010{\natexlab{b}})\citenamefont{{Yamazaki}, {Ichiki}, {Kajino}, and
  {Mathews}}}]{2010AIPC.1269...57Y}
\bibinfo{author}{\bibfnamefont{D.~G.} \bibnamefont{{Yamazaki}}},
  \bibinfo{author}{\bibfnamefont{K.}~\bibnamefont{{Ichiki}}},
  \bibinfo{author}{\bibfnamefont{T.}~\bibnamefont{{Kajino}}}, \bibnamefont{and}
  \bibinfo{author}{\bibfnamefont{G.~J.} \bibnamefont{{Mathews}}},
  \bibinfo{journal}{AIP Conference Proceedings}
  \textbf{\bibinfo{volume}{1269}}, \bibinfo{pages}{57}
  (\bibinfo{year}{2010}{\natexlab{b}}).

\bibitem[{\citenamefont{Brown and Crittenden}(2005)}]{Brown:2005kr}
\bibinfo{author}{\bibfnamefont{I.}~\bibnamefont{Brown}} \bibnamefont{and}
  \bibinfo{author}{\bibfnamefont{R.}~\bibnamefont{Crittenden}},
  \bibinfo{journal}{Phys. Rev.} \textbf{\bibinfo{volume}{D 72}},
  \bibinfo{pages}{063002} (\bibinfo{year}{2005}).

\bibitem[{\citenamefont{Seshadri and Subramanian}(2009)}]{Seshadri:2009sy}
\bibinfo{author}{\bibfnamefont{T.~R.} \bibnamefont{Seshadri}} \bibnamefont{and}
  \bibinfo{author}{\bibfnamefont{K.}~\bibnamefont{Subramanian}},
  \bibinfo{journal}{Phys. Rev. Lett.} \textbf{\bibinfo{volume}{103}},
  \bibinfo{pages}{081303} (\bibinfo{year}{2009}).

\bibitem[{\citenamefont{Caprini et~al.}(2009)\citenamefont{Caprini, Finelli,
  Paoletti, and Riotto}}]{Caprini:2009vk}
\bibinfo{author}{\bibfnamefont{C.}~\bibnamefont{Caprini}},
  \bibinfo{author}{\bibfnamefont{F.}~\bibnamefont{Finelli}},
  \bibinfo{author}{\bibfnamefont{D.}~\bibnamefont{Paoletti}}, \bibnamefont{and}
  \bibinfo{author}{\bibfnamefont{A.}~\bibnamefont{Riotto}},
  \bibinfo{journal}{JCAP} \textbf{\bibinfo{volume}{0906}}, \bibinfo{pages}{021}
  (\bibinfo{year}{2009}).

\bibitem[{\citenamefont{Kosowsky and Loeb}(1996)}]{Kosowsky:1996yc}
\bibinfo{author}{\bibfnamefont{A.}~\bibnamefont{Kosowsky}} \bibnamefont{and}
  \bibinfo{author}{\bibfnamefont{A.}~\bibnamefont{Loeb}},
  \bibinfo{journal}{Astrophys. J.} \textbf{\bibinfo{volume}{469}},
  \bibinfo{pages}{1} (\bibinfo{year}{1996}).

\bibitem[{\citenamefont{Kolatt}(1998)}]{Kolatt:1997xu}
\bibinfo{author}{\bibfnamefont{T.}~\bibnamefont{Kolatt}},
  \bibinfo{journal}{Astrophys. J.} \textbf{\bibinfo{volume}{495}},
  \bibinfo{pages}{564} (\bibinfo{year}{1998}).

\bibitem[{\citenamefont{Kosowsky et~al.}(2005)\citenamefont{Kosowsky,
  Kahniashvili, Lavrelashvili, and Ratra}}]{Kosowsky:2004zh}
\bibinfo{author}{\bibfnamefont{A.}~\bibnamefont{Kosowsky}},
  \bibinfo{author}{\bibfnamefont{T.}~\bibnamefont{Kahniashvili}},
  \bibinfo{author}{\bibfnamefont{G.}~\bibnamefont{Lavrelashvili}},
  \bibnamefont{and} \bibinfo{author}{\bibfnamefont{B.}~\bibnamefont{Ratra}},
  \bibinfo{journal}{Phys. Rev.} \textbf{\bibinfo{volume}{D 71}},
  \bibinfo{pages}{043006} (\bibinfo{year}{2005}).

\bibitem[{\citenamefont{Campanelli et~al.}(2004)\citenamefont{Campanelli,
  Dolgov, Giannotti, and Villante}}]{Campanelli:2004pm}
\bibinfo{author}{\bibfnamefont{L.}~\bibnamefont{Campanelli}},
  \bibinfo{author}{\bibfnamefont{A.~D.} \bibnamefont{Dolgov}},
  \bibinfo{author}{\bibfnamefont{M.}~\bibnamefont{Giannotti}},
  \bibnamefont{and} \bibinfo{author}{\bibfnamefont{F.~L.}
  \bibnamefont{Villante}}, \bibinfo{journal}{Astrophys. J.}
  \textbf{\bibinfo{volume}{616}}, \bibinfo{pages}{1} (\bibinfo{year}{2004}).

\bibitem[{\citenamefont{Chen et~al.}(2004)\citenamefont{Chen, Mukherjee,
  Kahniashvili, Ratra, and Wang}}]{Chen:2004nf}
\bibinfo{author}{\bibfnamefont{G.}~\bibnamefont{Chen}},
  \bibinfo{author}{\bibfnamefont{P.}~\bibnamefont{Mukherjee}},
  \bibinfo{author}{\bibfnamefont{T.}~\bibnamefont{Kahniashvili}},
  \bibinfo{author}{\bibfnamefont{B.}~\bibnamefont{Ratra}}, \bibnamefont{and}
  \bibinfo{author}{\bibfnamefont{Y.}~\bibnamefont{Wang}},
  \bibinfo{journal}{Astrophys. J.} \textbf{\bibinfo{volume}{611}},
  \bibinfo{pages}{655} (\bibinfo{year}{2004}).

\bibitem[{\citenamefont{Bernui and Hipolito-Ricaldi}(2008)}]{Bernui:2008ve}
\bibinfo{author}{\bibfnamefont{A.}~\bibnamefont{Bernui}} \bibnamefont{and}
  \bibinfo{author}{\bibfnamefont{W.~S.} \bibnamefont{Hipolito-Ricaldi}},
  \bibinfo{journal}{Mon. Not. Roy. Astron. Soc.}
  \textbf{\bibinfo{volume}{389}}, \bibinfo{pages}{1453} (\bibinfo{year}{2008}).

\bibitem[{\citenamefont{Sethi}(2003)}]{Sethi:2003vp}
\bibinfo{author}{\bibfnamefont{S.~K.} \bibnamefont{Sethi}},
  \bibinfo{journal}{Mon. Not. Roy. Astron. Soc.}
  \textbf{\bibinfo{volume}{342}}, \bibinfo{pages}{962} (\bibinfo{year}{2003}).

\bibitem[{\citenamefont{Sethi and Subramanian}(2005)}]{Sethi:2004pe}
\bibinfo{author}{\bibfnamefont{S.~K.} \bibnamefont{Sethi}} \bibnamefont{and}
  \bibinfo{author}{\bibfnamefont{K.}~\bibnamefont{Subramanian}},
  \bibinfo{journal}{Mon. Not. Roy. Astron. Soc.}
  \textbf{\bibinfo{volume}{356}}, \bibinfo{pages}{778} (\bibinfo{year}{2005}).

\bibitem[{\citenamefont{Jedamzik et~al.}(1998)\citenamefont{Jedamzik,
  Katalinic, and Olinto}}]{Jedamzik:1996wp}
\bibinfo{author}{\bibfnamefont{K.}~\bibnamefont{Jedamzik}},
  \bibinfo{author}{\bibfnamefont{V.}~\bibnamefont{Katalinic}},
  \bibnamefont{and} \bibinfo{author}{\bibfnamefont{A.~V.}
  \bibnamefont{Olinto}}, \bibinfo{journal}{Phys. Rev.}
  \textbf{\bibinfo{volume}{D 57}}, \bibinfo{pages}{3264}
  (\bibinfo{year}{1998}).

\bibitem[{\citenamefont{Subramanian and
  Barrow}(1998{\natexlab{b}})}]{Subramanian:1997gi}
\bibinfo{author}{\bibfnamefont{K.}~\bibnamefont{Subramanian}} \bibnamefont{and}
  \bibinfo{author}{\bibfnamefont{J.~D.} \bibnamefont{Barrow}},
  \bibinfo{journal}{Phys. Rev.} \textbf{\bibinfo{volume}{D 58}},
  \bibinfo{pages}{083502} (\bibinfo{year}{1998}{\natexlab{b}}).

\bibitem[{\citenamefont{Banerjee and Jedamzik}(2004)}]{Banerjee:2004df}
\bibinfo{author}{\bibfnamefont{R.}~\bibnamefont{Banerjee}} \bibnamefont{and}
  \bibinfo{author}{\bibfnamefont{K.}~\bibnamefont{Jedamzik}},
  \bibinfo{journal}{Phys. Rev.} \textbf{\bibinfo{volume}{D 70}},
  \bibinfo{pages}{123003} (\bibinfo{year}{2004}).

\bibitem[{\citenamefont{Bucher et~al.}(2000)\citenamefont{Bucher, Moodley, and
  Turok}}]{Bucher:1999re}
\bibinfo{author}{\bibfnamefont{M.}~\bibnamefont{Bucher}},
  \bibinfo{author}{\bibfnamefont{K.}~\bibnamefont{Moodley}}, \bibnamefont{and}
  \bibinfo{author}{\bibfnamefont{N.}~\bibnamefont{Turok}},
  \bibinfo{journal}{Phys. Rev.} \textbf{\bibinfo{volume}{D62}},
  \bibinfo{pages}{083508} (\bibinfo{year}{2000}).

\bibitem[{\citenamefont{Shaw and Lewis}(2010)}]{Shaw:2009nf}
\bibinfo{author}{\bibfnamefont{J.~R.} \bibnamefont{Shaw}} \bibnamefont{and}
  \bibinfo{author}{\bibfnamefont{A.}~\bibnamefont{Lewis}},
  \bibinfo{journal}{Phys. Rev.} \textbf{\bibinfo{volume}{D 81}},
  \bibinfo{pages}{043517} (\bibinfo{year}{2010}).

\bibitem[{\citenamefont{Yamamoto et~al.}(1997)\citenamefont{Yamamoto, Sugiyama,
  and Sato}}]{Yamamoto:1997qc}
\bibinfo{author}{\bibfnamefont{K.}~\bibnamefont{Yamamoto}},
  \bibinfo{author}{\bibfnamefont{N.}~\bibnamefont{Sugiyama}}, \bibnamefont{and}
  \bibinfo{author}{\bibfnamefont{H.}~\bibnamefont{Sato}},
  \bibinfo{journal}{ApJ.} \textbf{\bibinfo{volume}{501}}, \bibinfo{pages}{442}
  (\bibinfo{year}{1997}).

\bibitem[{\citenamefont{Seshadri and Subramanian}(2001)}]{Seshadri:2000ky}
\bibinfo{author}{\bibfnamefont{T.~R.} \bibnamefont{Seshadri}} \bibnamefont{and}
  \bibinfo{author}{\bibfnamefont{K.}~\bibnamefont{Subramanian}},
  \bibinfo{journal}{Phys. Rev. Lett.} \textbf{\bibinfo{volume}{87}},
  \bibinfo{pages}{101301} (\bibinfo{year}{2001}).

\bibitem[{\citenamefont{Starobinskii}(1979)}]{1979ZhPmR..30..719S}
\bibinfo{author}{\bibfnamefont{A.~A.} \bibnamefont{Starobinskii}},
  \bibinfo{journal}{ZhETF Pis ma Redaktsiiu} \textbf{\bibinfo{volume}{30}},
  \bibinfo{pages}{719} (\bibinfo{year}{1979}).

\bibitem[{\citenamefont{Rubakov et~al.}(1982)\citenamefont{Rubakov, Sazhin, and
  Veryaskin}}]{1982PhLB..115..189R}
\bibinfo{author}{\bibfnamefont{V.~A.} \bibnamefont{Rubakov}},
  \bibinfo{author}{\bibfnamefont{M.~V.} \bibnamefont{Sazhin}},
  \bibnamefont{and} \bibinfo{author}{\bibfnamefont{A.~V.}
  \bibnamefont{Veryaskin}}, \bibinfo{journal}{Physics Letters B}
  \textbf{\bibinfo{volume}{115}}, \bibinfo{pages}{189} (\bibinfo{year}{1982}).

\bibitem[{\citenamefont{Polnarev}(1985)}]{1985SvA....29..607P}
\bibinfo{author}{\bibfnamefont{A.~G.} \bibnamefont{Polnarev}},
  \bibinfo{journal}{Soviet Astronomy} \textbf{\bibinfo{volume}{29}},
  \bibinfo{pages}{607} (\bibinfo{year}{1985}).

\bibitem[{\citenamefont{Pritchard and Kamionkowski}(2005)}]{Pritchard:2004qp}
\bibinfo{author}{\bibfnamefont{J.~R.} \bibnamefont{Pritchard}}
  \bibnamefont{and}
  \bibinfo{author}{\bibfnamefont{M.}~\bibnamefont{Kamionkowski}},
  \bibinfo{journal}{Ann. Phys.} \textbf{\bibinfo{volume}{318}},
  \bibinfo{pages}{2} (\bibinfo{year}{2005}).

\bibitem[{\citenamefont{Peebles}(1980)}]{Peebles:1980booka}
\bibinfo{author}{\bibfnamefont{P.~J.~E.} \bibnamefont{Peebles}},
  \emph{\bibinfo{title}{The Large-Scale Structure of the Universe}}
  (\bibinfo{publisher}{Princeton University Press}, \bibinfo{year}{1980}).

\bibitem[{\citenamefont{Spergel et~al.}(2006)}]{Spergel:2006hy}
\bibinfo{author}{\bibfnamefont{D.~N.} \bibnamefont{Spergel}}
  \bibnamefont{et~al.} (\bibinfo{year}{2006}), \eprint{astro-ph/0603449}.

\bibitem[{\citenamefont{Hinshaw et~al.}(2006)}]{Hinshaw:2006ia}
\bibinfo{author}{\bibfnamefont{G.}~\bibnamefont{Hinshaw}} \bibnamefont{et~al.}
  (\bibinfo{year}{2006}), \eprint{astro-ph/0603451}.

\bibitem[{\citenamefont{Page et~al.}(2006)}]{Page:2006hz}
\bibinfo{author}{\bibfnamefont{L.}~\bibnamefont{Page}} \bibnamefont{et~al.}
  (\bibinfo{year}{2006}), \eprint{astro-ph/0603450}.

\bibitem[{\citenamefont{Lewis and Bridle}(2002)}]{Lewis:2002ah}
\bibinfo{author}{\bibfnamefont{A.}~\bibnamefont{Lewis}} \bibnamefont{and}
  \bibinfo{author}{\bibfnamefont{S.}~\bibnamefont{Bridle}},
  \bibinfo{journal}{Phys. Rev.} \textbf{\bibinfo{volume}{D 66}},
  \bibinfo{pages}{103511} (\bibinfo{year}{2002}).

\bibitem[{\citenamefont{Cole et~al.}(2005)}]{Cole:2005sx}
\bibinfo{author}{\bibfnamefont{S.}~\bibnamefont{Cole}} \bibnamefont{et~al.}
  (\bibinfo{collaboration}{The 2dFGRS}), \bibinfo{journal}{Mon. Not. Roy.
  Astron. Soc.} \textbf{\bibinfo{volume}{362}}, \bibinfo{pages}{505}
  (\bibinfo{year}{2005}).

\bibitem[{\citenamefont{Tegmark et~al.}(2006)}]{Tegmark:2006az}
\bibinfo{author}{\bibfnamefont{M.}~\bibnamefont{Tegmark}} \bibnamefont{et~al.}
  (\bibinfo{collaboration}{SDSS}), \bibinfo{journal}{Phys. Rev.}
  \textbf{\bibinfo{volume}{D74}}, \bibinfo{pages}{123507}
  (\bibinfo{year}{2006}).

\bibitem[{\citenamefont{Rozo et~al.}(2007)}]{Rozo:2007yt}
\bibinfo{author}{\bibfnamefont{E.}~\bibnamefont{Rozo}} \bibnamefont{et~al.}
  (\bibinfo{year}{2007}), \eprint{astro-ph/0703571}.

\bibitem[{\citenamefont{Ross et~al.}(2008)\citenamefont{Ross, Brunner, and
  Myers}}]{Ross:2008ze}
\bibinfo{author}{\bibfnamefont{A.~J.} \bibnamefont{Ross}},
  \bibinfo{author}{\bibfnamefont{R.~J.} \bibnamefont{Brunner}},
  \bibnamefont{and} \bibinfo{author}{\bibfnamefont{A.~D.} \bibnamefont{Myers}}
  (\bibinfo{year}{2008}), \eprint{0804.3325}.

\bibitem[{\citenamefont{Dunkley et~al.}(2009)}]{Dunkley:2008ie}
\bibinfo{author}{\bibfnamefont{J.}~\bibnamefont{Dunkley}} \bibnamefont{et~al.}
  (\bibinfo{collaboration}{WMAP}), \bibinfo{journal}{Astrophys. J. Suppl.}
  \textbf{\bibinfo{volume}{180}}, \bibinfo{pages}{306} (\bibinfo{year}{2009}).

\bibitem[{\citenamefont{{Hinshaw} et~al.}(2009)\citenamefont{{Hinshaw},
  {Weiland}, {Hill}, {Odegard}, {Larson}, {Bennett}, {Dunkley}, {Gold},
  {Greason}, {Jarosik} et~al.}}]{Hinshaw:2008kr}
\bibinfo{author}{\bibfnamefont{G.}~\bibnamefont{{Hinshaw}}},
  \bibinfo{author}{\bibfnamefont{J.~L.} \bibnamefont{{Weiland}}},
  \bibinfo{author}{\bibfnamefont{R.~S.} \bibnamefont{{Hill}}},
  \bibinfo{author}{\bibfnamefont{N.}~\bibnamefont{{Odegard}}},
  \bibinfo{author}{\bibfnamefont{D.}~\bibnamefont{{Larson}}},
  \bibinfo{author}{\bibfnamefont{C.~L.} \bibnamefont{{Bennett}}},
  \bibinfo{author}{\bibfnamefont{J.}~\bibnamefont{{Dunkley}}},
  \bibinfo{author}{\bibfnamefont{B.}~\bibnamefont{{Gold}}},
  \bibinfo{author}{\bibfnamefont{M.~R.} \bibnamefont{{Greason}}},
  \bibinfo{author}{\bibfnamefont{N.}~\bibnamefont{{Jarosik}}},
  \bibnamefont{et~al.}, \bibinfo{journal}{Astrophys. J}
  \textbf{\bibinfo{volume}{180}}, \bibinfo{pages}{225} (\bibinfo{year}{2009}).

\bibitem[{\citenamefont{Kuo et~al.}(2007)\citenamefont{Kuo, Ade, Bock, Bond,
  Contaldi, Daub, Goldstein, Holzapfel, Lange, Lueker et~al.}}]{Kuo:2006ya}
\bibinfo{author}{\bibfnamefont{C.~L.} \bibnamefont{Kuo}},
  \bibinfo{author}{\bibfnamefont{P.~A.~R.} \bibnamefont{Ade}},
  \bibinfo{author}{\bibfnamefont{J.~J.} \bibnamefont{Bock}},
  \bibinfo{author}{\bibfnamefont{J.~R.} \bibnamefont{Bond}},
  \bibinfo{author}{\bibfnamefont{C.~R.} \bibnamefont{Contaldi}},
  \bibinfo{author}{\bibfnamefont{M.~D.} \bibnamefont{Daub}},
  \bibinfo{author}{\bibfnamefont{J.~H.} \bibnamefont{Goldstein}},
  \bibinfo{author}{\bibfnamefont{W.~L.} \bibnamefont{Holzapfel}},
  \bibinfo{author}{\bibfnamefont{A.~E.} \bibnamefont{Lange}},
  \bibinfo{author}{\bibfnamefont{M.}~\bibnamefont{Lueker}},
  \bibnamefont{et~al.}, \bibinfo{journal}{Astrophys. J.}
  \textbf{\bibinfo{volume}{664}}, \bibinfo{pages}{687} (\bibinfo{year}{2007}).

\bibitem[{\citenamefont{{Sievers} et~al.}(2007)\citenamefont{{Sievers},
  {Achermann}, {Bond}, {Bronfman}, {Bustos}, {Contaldi}, {Dickinson},
  {Ferreira}, {Jones}, {Lewis} et~al.}}]{Sievers:2005gj}
\bibinfo{author}{\bibfnamefont{J.~L.} \bibnamefont{{Sievers}}},
  \bibinfo{author}{\bibfnamefont{C.}~\bibnamefont{{Achermann}}},
  \bibinfo{author}{\bibfnamefont{J.~R.} \bibnamefont{{Bond}}},
  \bibinfo{author}{\bibfnamefont{L.}~\bibnamefont{{Bronfman}}},
  \bibinfo{author}{\bibfnamefont{R.}~\bibnamefont{{Bustos}}},
  \bibinfo{author}{\bibfnamefont{C.~R.} \bibnamefont{{Contaldi}}},
  \bibinfo{author}{\bibfnamefont{C.}~\bibnamefont{{Dickinson}}},
  \bibinfo{author}{\bibfnamefont{P.~G.} \bibnamefont{{Ferreira}}},
  \bibinfo{author}{\bibfnamefont{M.~E.} \bibnamefont{{Jones}}},
  \bibinfo{author}{\bibfnamefont{A.~M.} \bibnamefont{{Lewis}}},
  \bibnamefont{et~al.}, \bibinfo{journal}{Astrophys. J.}
  \textbf{\bibinfo{volume}{660}}, \bibinfo{pages}{976} (\bibinfo{year}{2007}).

\bibitem[{\citenamefont{Jones et~al.}(2006)}]{Jones:2005yb}
\bibinfo{author}{\bibfnamefont{W.~C.} \bibnamefont{Jones}}
  \bibnamefont{et~al.}, \bibinfo{journal}{Astrophys. J.}
  \textbf{\bibinfo{volume}{647}}, \bibinfo{pages}{823} (\bibinfo{year}{2006}).

\bibitem[{\citenamefont{Caprini and Durrer}(2001)}]{Caprini:2001nb}
\bibinfo{author}{\bibfnamefont{C.}~\bibnamefont{Caprini}} \bibnamefont{and}
  \bibinfo{author}{\bibfnamefont{R.}~\bibnamefont{Durrer}},
  \bibinfo{journal}{Phys. Rev.} \textbf{\bibinfo{volume}{D 65}},
  \bibinfo{pages}{023517} (\bibinfo{year}{2001}).

\bibitem[{\citenamefont{Durrer and Caprini}(2003)}]{Durrer:2003ja}
\bibinfo{author}{\bibfnamefont{R.}~\bibnamefont{Durrer}} \bibnamefont{and}
  \bibinfo{author}{\bibfnamefont{C.}~\bibnamefont{Caprini}},
  \bibinfo{journal}{JCAP} \textbf{\bibinfo{volume}{0311}}, \bibinfo{pages}{010}
  (\bibinfo{year}{2003}), \eprint{astro-ph/0305059}.

\bibitem[{\citenamefont{Kojima and Ichiki}(2009)}]{Kojima:2009ms}
\bibinfo{author}{\bibfnamefont{K.}~\bibnamefont{Kojima}} \bibnamefont{and}
  \bibinfo{author}{\bibfnamefont{K.}~\bibnamefont{Ichiki}}
  (\bibinfo{year}{2009}), \eprint{arXiv:0902.1367}.

\bibitem[{\citenamefont{Durrer et~al.}(2000)\citenamefont{Durrer, Ferreira, and
  Kahniashvili}}]{Durrer:1999bk}
\bibinfo{author}{\bibfnamefont{R.}~\bibnamefont{Durrer}},
  \bibinfo{author}{\bibfnamefont{P.~G.} \bibnamefont{Ferreira}},
  \bibnamefont{and}
  \bibinfo{author}{\bibfnamefont{T.}~\bibnamefont{Kahniashvili}},
  \bibinfo{journal}{Phys. Rev.} \textbf{\bibinfo{volume}{D 61}},
  \bibinfo{pages}{043001} (\bibinfo{year}{2000}).

\bibitem[{\citenamefont{Yoshida et~al.}(2003)\citenamefont{Yoshida, Sugiyama,
  and Hernquist}}]{Yoshida:2003sy}
\bibinfo{author}{\bibfnamefont{N.}~\bibnamefont{Yoshida}},
  \bibinfo{author}{\bibfnamefont{N.}~\bibnamefont{Sugiyama}}, \bibnamefont{and}
  \bibinfo{author}{\bibfnamefont{L.}~\bibnamefont{Hernquist}},
  \bibinfo{journal}{Mon. Not. Roy. Astron. Soc.}
  \textbf{\bibinfo{volume}{344}}, \bibinfo{pages}{481} (\bibinfo{year}{2003}).

\end{thebibliography}

\end{document}